# 3-D Atomic Mapping of Interfacial Roughness and its Spatial Correlation Length in sub-10 nm Superlattices

Samik Mukherjee, Anis Attiaoui, Matthias Bauer, and Oussama Moutanabbir*

Dr. Samik Mukherjee, Anis Attiaoui, and Prof. Oussama Moutanabbir
Department of Engineering Physics, École Polytechnique de Montréal, C. P. 6079, Succ. Centre-Ville, Montreal, Québec H3C 3A7, Canada
E-mail: oussama.moutanabbie@polymtl.ca

Dr. Matthias Bauer
Applied Materials Inc., 974 E. Arques Avenue, Sunnyvale, CA 94085, USA

**Abstract**

The interfacial abruptness and uniformity in heterostructures are critical to control their electronic and optical properties. With this perspective, this work demonstrates the 3-D atomistic-level mapping of the roughness and uniformity of buried epitaxial interfaces in Si/SiGe superlattices with a layer thickness in the 1.5-7.5 nm range. Herein, 3-D atom-by-atom maps were acquired and processed to generate iso-concentration surfaces highlighting local fluctuations in content at each interface. These generated surfaces were subsequently utilized to map the interfacial roughness and its spatial correlation length. The analysis revealed that the root mean squared roughness of the buried interfaces in the investigated superlattices is sensitive to the growth temperature with a value varying from ∼0.2 nm (±13%) to ∼0.3 nm (±11.5%) in the temperature range of 500-650 °C. The estimated horizontal correlation lengths were found to be 8.1 nm (±5.8%) at 650 °C and 10.1 nm (±6.2%) at 500 °C. Additionally, reducing the growth temperature was found to improve the interfacial abruptness, with 30 % smaller interfacial width is obtained at 500 °C. This behavior is attributed to the thermally activated atomic exchange at the surface during the heteroepitaxy. Finally, by testing different optical models with increasing levels of interfacial complexity, it is demonstrated that the observed atomic-level roughening at the interface must be accounted for to accurately describe the optical response of Si/SiGe heterostructures.

Heterostructures have been a rich platform to engineer a variety of low-dimensional structures and devices.[1–6] In such systems, the nature of the interfaces is a crucial factor that ultimately defines their basic properties and performance. For instance, it is well known that several terahertz to infrared sources and detectors operate on the basis of intersubband transitions across semiconductor quantum wells, wherein the linewidth of these transitions depends strongly on the interface roughness.[7,8] This becomes more prominent in quantum cascade structures where the electronic states are spread out over several quantum wells and encompass several interfaces whose roughness is not correlated.[4,7] The spatial correlation of the vertical height distribution of buried epitaxial interfaces is important parameter to evaluate the performance of quantum cascade structures because it underlies crucial information required to evaluate the scattering matrix.[9] In fact, the interface roughness in a cascaded structure induces intersubband scattering between electronic states when the correlation length matches the inverse of the momentum needed for the process. However, despite their importance, direct measurements of the horizontal correlation length for the buried interfaces are still conspicuously missing in literature. As a matter of fact, this correlation length is currently used as a fitting parameter to theoretical models.[10] Additionally, the precise knowledge of the roughness of buried interfaces has also become increasingly critical in silicon (Si) gate-all-around designs recently introduced for the 7 nm technology node and beyond.[11] These architectures are based on Si/SiGe superlattices (SLs) where selective wet-etching of the SiGe layers is used to release the Si layers and form vertically stacked Si nanosheets. The Si/SiGe interfacial width and irregularities before the etching are expected to determine the roughness of Si nanosheets and hence the extent of charge carrier surface scattering and the overall performance of the final device. The control of the interfacial abruptness and the thickness uniformity has also implications in the development of Si-based direct bandgap semiconductor materials by superimposing a periodic SL potential onto the crystal lattice potential.[12] A periodic sequence of a few atomic planes of Si and Ge leads to a new larger lattice constant in one direction and, consequently, the Brillouin zone is reduced along this axis. A proper choice of the SL period length results in a Brillouin zone folding such that initially indirect conduction band minima are shifted back to the center of the reduced Brillouin zone, giving rise to direct electronic transitions. Inevitably, the implementation of these photonic structures requires a meticulous control of the roughness and abruptness at each interface.

The elements above highlight the importance of understanding the interfacial properties of epitaxial multilayer structures. With this perspective, this work describes a method to achieve a 3-D atomistic-level mapping of the roughness and uniformity of buried epitaxial interfaces in a variety of $(Si)_m/(Si_{1-x}Ge_x)_m$ ('m' being the periodicity) SLs. The growth of the samples are described in the experimental method section. The samples are labeled based on their periodicities, 'm'. For example, the sample with 16 periods (m = 16) is named S-16, and so on. In addition to S-16, the other investigated samples are S-12, S-6, and S-3. S-16 and S-12 were grown at 650 ℃, S-6 at 600 ℃, and S-3 at 500 ℃. The mean Ge concentration of the $Si_{1-x}Ge_x$ layers within the SLs is in the ~25 to ~30 at.% range, and their thickness varies from ~40 nm to ~60 nm (Table 1). Herein, cross-sectional transmission electron microscopy (XTEM) and three-dimensional (3-D) atom probe tomography (APT) are combined to elucidate the properties of heteroepitaxial interfaces (see details in experimental method). Spectroscopic ellipsometry (SE) models at increasing levels of complexity were developed to evaluate the impact of the interfacial sharpness on the optical response of the investigated heterostructures (see details in experimental method). **Figure** 1(a) shows the XTEM images of sample S-16. The XTEM image of other investigated SLs are shown in **Figure** S1 in the supporting information (SI). $Si_{1-x}Ge_x$ layers appear brighter than Si layers in the XTEM images in **Figure** 1(a). The insets are zoom-in XTEM images of different regions in the SL.

The $Si_{1-x}Ge_x$/Si interfaces in the XTEM images appear to be coherent without the presence of any extended defects. The mean thickness of each $Si_{1-x}Ge_x$ and Si layer in different samples is presented in Table 1. **Figure** 1(b) shows the 3-D atom-by-atom reconstruction of SL S-16. Only one representative 3-D reconstruction is shown. Out of all the samples studied, S-16 possesses the highest number of interfaces and therefore provides the larger statistics on any interfacial parameter that is being measured. Hence, the APT investigations of the interface properties have been demonstrated in the following sections using this SL, but the same methodology was applied to investigate the interfacial properties in all the other SLs. The evaporation of atoms being the most uniform from the center of an APT tip,[13] only atoms from within the black cylinder (diameter 30 nm) at the center of the 3-D reconstruction in **Figure** 1(b) were extracted and the corresponding 3-D reconstruction is shown in **Figure** 1(c)-top. Note that as the 3-D reconstruction in APT can often be subjected to artifacts as far as the length scales of features are concerned. Thus, the correlation with XTEM data has been followed to optimize the APT reconstruction parameters, as shown in **Figure** 1(c)-bottom. The APT reconstruction of all samples investigated in this work was done iteratively until the layer thickness of the 3-D APT reconstruction matches (with 5.0 % tolerance) that obtained from the corresponding XTEM image. Furthermore, 2-3 tips from each sample were analyzed in APT to gather better statistics and verify the sample uniformity.

**Figure** S2 shows the full 1-D concentration profile of Si and Ge across all the 16 periods of SL S-16, recorded from both APT and electron energy loss spectroscopy (EELS) data. The estimated average Ge concentration within the $Si_{1-x}Ge_x$ layers is ~24.5 at.% (±2.0%). The bottom most $Si_{1-x}Ge_x$ layer, which grew thicker than the rest of the layers, possesses a relatively higher Ge concentration of ~30.0 at.%. The mean Ge contents of the $Si_{1-x}Ge_x$ layers, as extracted from APT, are given in Table 1. **Figure** 1(d) shows the buried $Si/Si_{1-x}Ge_x$

interfaces, drawn as iso-concentration surfaces within the 3D reconstruction (shown in **Figure** 1(c)-top) defined at 50 % of the mean Ge concentration in the $Si_{1-x}Ge_x$ layers. The interface number increases in the direction of APT evaporation sequence. In the next section, the roughness and spatial correlation have been extracted for these interfaces, defined as iso-concentration surface. Clearly, the top 3-4 interfaces (enclosed within the partially transparent black rectangular box in **Figure** 1(d)) were slightly damaged during the FIB preparation and have been omitted from the analysis. The method implemented to extract the root mean squared (RMS) roughness and horizontal correlation length is explained by randomly selecting interface number 24 (marked by the green rectangle in **Figure** 1(d)) as an example. First, the interface properties were exported and the vertical height (z) at every point over the interface was evaluated and plotted as a color-coded image, as shown in the inset of **Figure** 2(a). The height-height correlation function $H(\tau)$ is the squared difference in height of two points $(x, y)$ and $(x', y')$ separated by a distance $\tau$ and is given by $H(\tau) = \langle |z(x,y) - z(x',y')|^2 \rangle$, where $z(x, y)$ is the height of the interface at the position $(x, y)$, relative to a mean plane and $\tau = [(x - x')^2 + (y - y')^2]^{1/2}$.[14,15] For a pixelated color-coded image like the one in the inset of **Figure** 2(a), the following equation was used to calculate the correlation function:

$$H(\tau) = \frac{1}{N'(M' - m')} \sum_{l'=1}^{N'} \sum_{n'=1}^{M'-m'} \left( z_{n'+m',l'} - z_{n',l'} \right)^2 \qquad (1)$$

where $M'$ and $m'$ are respectively the total number of pixels and the separation between two pixels on the image during a line scan along an arbitrarily chosen axis, and $N'$ is the total number of scan lines required to encompass the whole interface. The pixel size in the color-coded images are the grid parameter (1.0 nm × 1.0 nm × 1.0 nm voxel size) used during the 3-D reconstruction. The evolution of H(τ) as a function of τ for interface number 24 (from top) of sample S-16 is shown in **Figure** 2(a).

Phenomenologically, the height-height correlation can be fitted by the function $H_{fit}(\tau) = 2\sigma^2 \left[1 - e^{-(\tau/\xi)^{2\alpha}}\right]$, where σ is the RMS roughness, ξ is the horizontal correlation length, and α is called the Hurst parameter.[14,15] The data in **Figure** 2(a) was fitted with the function $H_{fit}(\tau)$ (red line) in order to estimate the parameters σ and ξ. It is clear from **Figure** 2(a) that H(τ) exhibits two distinct regimes. First, a linear increase where H(τ) strongly depends on the value of τ, implying a strong correlation between any two points situated at (x, y) and (x′, y′). Second, the regime where H(τ) becomes independent of τ, implying no correlation between any two points situated at (x, y) and (x′, y′). The point of crossover from the first regime to the second, where the correlation function starts to flatten out is the one that defines the horizontal correlation length ξ. For interface number 24 of S-16, σ of ∼0.2 nm and ξ of ∼8.4 nm was estimated from the fit. A similar example using a randomly selected interface of sample S-3 is shown in **Figure** S3. **Figure** 2(b)-(c) show the values of σ and ξ of all the interfaces over two different data sets of S-16, respectively. From the data, no meaningful differences can be observed between the interfaces at $Si_{1-x}Ge_x \rightarrow Si$ transitions and those at the reverse transition. The mean values (approximated to the first decimal place) of σ and ξ was found to be ∼0.3 nm (±11.5%) and ∼8.1 nm (±5.8%), respectively. The uncertainty here is the standard deviation over all the interfaces and over two different APT data sets. As shown in **Figure** S4, for sample S-6 the mean values of σ and ξ was found to be ∼0.3 nm (±12.2%) and ∼8.4 nm (±6.0%), respectively, similar (within the uncertainty) to that of S-16. However, the mean values of σ and ξ for sample S-3 were found to be ∼0.2 nm (±13.0%) and ∼10.1 nm (±6.2%), respectively, as evident from **Figure** S5. A graphical representation of the evolution of the mean values of σ and ξ of the SLs as a function of the growth temperature is presented in **Figure** 2(d).

A segment of the 1-D concentration profile (at a fixed bin width of 0.2 nm) of the last 7 periods of sample S-16 is shown in **Figure** 3(a). The profile in **Figure** 3(a) was collected from within the white cylinder, placed at the center of 3-D APT reconstruction in the inset of **Figure** 1(c)-bottom. The error bars in the concentration profiles in **Figure** 3(a) are IVAS-generated one-sigma statistical error, the magnitudes of which are $\sqrt{C_i(1-C_i)/N}$, where $C_i$ is atomic fraction of the i[th] element within a bin and N is the total number of atoms within a bin.[16] The inset of **Figure** 3(a) shows the rising (top) and falling (bottom) Ge concentration profiles, zoomed into the 12[th] and the 9[th] interfaces from the bottom of the SL. In this work, rising (Si→ $Si_{1-x}Ge_x$ transition) and falling ($Si_{1-x}Ge_x$ →Si transition) are denoted relative to the direction of evaporation in APT, which is opposite to the growth direction. The raw data was fitted using the sigmoid function:

$$c(x) = c_0 + d_0 / \left[1 + e^{-\frac{(x_0 \pm x)}{\mathcal{L}}}\right] \qquad (2)$$

where $c_0$ is a vertical positioning parameter, $x_0$ is a horizontal positioning parameter, $d_0$ is a scaling parameter and the parameter $\mathcal{L}$ determines the value of the interface width.[17] For the two interfaces shown in the inset of **Figure** 3(a), interface widths were estimated to be 0.95 nm and 0.96 nm. The width of all the interfaces (spread over 2 different data sets) of sample S-16, as extracted from APT is shown in **Figure** 3(b). Clearly, there is no difference between Si→ $Si_{1-x}Ge_x$ and $Si_{1-x}Ge_x$ → Si transitions. Overall, the interface widths are randomly distributed with a mean value of ~1.0 nm (±2.8%). The 1D Ge concentration profile (binned at 0.2 nm) of sample S-6 is shown in **Figure** S6(a). The average Ge concentration within the $Si_{1-x}Ge_x$ layers of sample S-6 is ~24.4 at.% (±2.0%). **Figure** S6(b) shows that the mean value of the interface widths in this SL is ~0.9 nm (±3.0%), close to that of SL S-16. However, SL S-3 exhibits a different behavior. The comparison of the 1-D concentration profiles from APT and EELS for sample S-3 is shown in **Figure** S7(a) and (b), respectively. The average Ge concentration within the $Si_{1-x}Ge_x$ layers of sample S-3 is about ~30.7 at.% (±1.5%). **Figure** S7(c) shows that the mean value of the interface width of sample S-3 (over 3 different data sets) is ~0.7 nm (±3.4%), about 30% smaller than that of S-16. A graphical comparison, showing the evolution of the interface width with the growth temperature of the samples, is displayed in **Figure** 3(c). Note that the interface width extracted from EELS data is always larger than that extracted from APT by about 10% for the thickest period SLs (S-3) and by about 26% for the thinnest SLs (S-16). For the thinnest SL layers, the interface widths obtained from EELS can be limited by the spot size of the electron beam.

The mean value of σ for S-3 is about 34% smaller while the mean value of ξ is 24% larger as compared to that of S-16. The interface properties of sample S-6 (grown at 600 °C) are quite similar to that of sample S-16. As far as the interface widths are concerned, the data suggests that there is no Ge segregation during the Si overlayer growth atop the $Si_{1-x}Ge_x$ layers in any of the samples. The phenomenon of Ge segregation has long been thought to be one of the main processes limiting the realization of sharp interfaces with identical widths for both $Si_{1-x}Ge_x$ → Si transition and the reverse transition.[18,19] The fact that the width and roughness of the $Si_{1-x}Ge_x$ → Si transitions are comparable to that of the reverse transitions indicates that Ge segregation is suppressed during the growth of SLs. Note that the rate of Ge surface segregation is often reduced when the growing surface is covered by some surfactant atoms like H or chlorine.[20] The surface H that is produced by the carrier gas as well as from the dissociation of disilane and germane on the growth front seems to be effective in preventing the Ge segregation. Bulk diffusion can be reasonably assumed to play an insignificant role, since it is an energetically unfavorable process.[21,22] The interface width for the samples investigated here might simply be determined by an atomic exchange process between the sub-surface atoms and ad-atoms on the growing surface. This is a kinetically controlled surface phenomenon, with activation energies of a fraction of an eV.[23] In the absence of bulk diffusion, the atomic exchange in a surface layer halts when the subsequent growing layer sweeps across the entire surface. The analysis shows that the interface roughness is more significant for a wider interface as compared to a relatively sharper interface. The buried interfaces are defined as iso-concentration surfaces, which were created in the first place by fitting the 3-D atomic distribution (within each voxel) by polygons. More intermixing across the interfaces consequently leads to a larger uncertainty in placing polygons within each voxel, precisely at the predefined Ge concentration. The interfaces, defined as iso-concentration surface, which are nothing but a combination of all these polygons, turn out to be rougher when compared to an interface with relatively less intermixing. In the following, the effects of this atomic-level roughening on the optical properties of the investigated SLs are discussed.

SE studies were first carried out to independently assess the thickness of the buried interfaces. The recorded optical response is shown here to provide a quantitative measure of the interfacial broadening in the investigated SLs. To build any ellipsometry model, an initial estimate of the thicknesses and composition of each layer is required. To this end, symmetric (004) and asymmetric (224) HRXRD spectra (not shown here) were measured for all SLs and fitted using standard dynamical simulations. HRXRD was used to ensure that the SE investigations of the SL interface widths are completely independent of any inputs from XTEM and APT. Table 1 displays a summary of the measurements conducted on different SLs. Note that SE-based analysis of the thickness shows a relative fluctuation between 24% and 46% when compared to XTEM. The full details of the considered optical models are provided in section 3.1, 3.2 and **Figure** S8 of the SI. **Figure** S8(a) shows a layer-by-layer optical model ($M_X$) where the interfaces are simply omitted (i.e., no interfacial broadening). In contrast, **Figure** S8(b) illustrates a model where the interfacial layers have been incorporated ($M_{int}$). In brief, in the $M_{int}$ model, the initial layer thicknesses $d_{Si}^{(i)}$ and $d_{SiGe}^{(i)}$ were obtained from HRXRD. All the optical models take into account the Bruggeman effective medium approximation (B-EMA).[24–28] The approximation combines the dielectric functions of two adjacent layers to form the dielectric constant of the interface. In the model labelled $M_{int}^{1-EMA}$, the dielectric constant of an interface was taken as a combination of the dielectric constants of Si and Ge. However, for $M_{int}^{2-EMA}$, a combination of the dielectric constant of $Si_{1-x}Ge_x$ and Si was used. In the $M_{int}^{1-EMA}$ model, EMA % represents the average Ge content within the host material Si. Third, a new parametric graded interfacial alloy model ($M_{int}^{\sigma}$) was introduced where the Si content inside the interface layer is graded (described by equation (2)), and the floating parameters are the scaling parameter $d_0$ (the average Ge concentration at the interface) and the interfacial width (labeled $d_{int}^{(i)}$ for the i$^{th}$ interface, in the optical models). An iterative process was developed to extract the desired set of parameters. The first iteration involved varying the layer thicknesses ($d_{SiGe}^{(i)}$ and $d_{Si}^{(i)}$) and keeping fixed the interface width ($d_{int}^{(i)}$). In the second iteration, each layer thickness was fixed and $d_{int}^{(i)}$ was varied. This iterative process is repeated until the mean-squared error (MSE, see sections 3.1 and 3.2 for more details) between two consecutive steps is smaller than a set tolerance value of 10$^{-3}$ and the gradient of $\Delta_{Err}^{S-m}$ (equation (3)) is minimized. Knowing the optical properties of Si and $Si_{1-x}Ge_x$ thin layers is also required to implement the optical models. To this end, pseudomorphic $Si_{1-x}Ge_x$ layers with Ge content below 54% and a thickness between 19 and 33 nm were used to extract the optical properties of $Si_{1-x}Ge_x$ layers in the SLs.[29] Those of Si layers were evaluated from a reference sample consisting of 12 nm-thick silicon-on-insulator (SOI), which was characterized for an angle of incidence (AOI) between 60° and 85° (see **Figure** S9).

**Figures** 4(a) and (b) exhibit the measured and fitted ellipsometric parameters $\Psi$ and $\Delta$, respectively, for S-6 at an AOI between 20° and 85°. Note that the same analysis was conducted for other SLs as well. Overlaid in the same figure are the fit to the experimental parameters using the four different optical models described above. The full red line corresponds to the $M_X$ model, whereas the dashed blue, purple and black lines represent the $M_{int}^{1-EMA}$, $M_{int}^{2-EMA}$, and $M_{int}^{\sigma}$ models, respectively. The difference between the optical models can be clearly seen in the inset of **Figure** 4(b). When analyzing S-6, the fit was found to be excellent for all spectra with an MSE between 3.3 and 5.5 for the proposed models. Nonetheless, a closer inspection of the fitted spectra reveals a different observation. Indeed, the inset in **Figure** 4(b) shows, in loglog scale, the SE parameter $\Delta$ at the highest incidence angle (80° and 85°) for a spectral range between 1 and 5 eV where the most features in the dataset are noticed. Above 2.6 eV, almost all the models accurately reproduces the experimental measurement, whereas below 2.6 eV, only the $M_{int}^{\sigma}$ model (dashed-black line) mirrors the measured dataset the best. This is further confirmed by comparing the obtained MSE of the four models, where the smallest MSE of 2.3 corresponds to $M_{int}^{\sigma}$. **Figure** 5(a) displays the variation of the average thickness relative error $\Delta_{Err}^{S-m}$ (%) as a function of the number of periods m for the $M_X$ model. $\Delta_{Err}^{S-m}$ is defined as follows:

$$\Delta_{Err}^{S-m} = 100 \times \sum_{i=1}^{m} \left| \frac{d_{SE}^{(i)} - d_{XTEM}^{(i)}}{d_{XTEM}^{(i)}} \right| \qquad (3)$$

where $d_{SE}^{(i)}$ and $d_{XTEM}^{(i)}$ represents the SE- and the XTEM-extracted thickness of the $i^{th}$ layer, respectively. The small MSE values need to be cross-correlated to a small average relative error ($\Delta_{Err}^{S-m}$) to evaluate the accuracy of the model. In fact, even though the MSE values are relatively smaller for m ≤ 12, which usually indicates a good fit quality and an accurate optical model, the $\Delta_{Err}^{S-m}$ is in general high for all the SL samples. This is a clear indication that the $M_X$ model overestimates the layer thicknesses in all investigated SLs, justifying the need for a more elaborate optical model. The first column of Table 2 shows in detail the systematic decrease in the error of the estimated layer thickness relative to XTEM measurements. For instance, a reduction of $\Delta_{Err}^{S-12}$ from 72% to 51% is observed by adding the additional EMA interfacial layer for S-12. The same iterative routine was also used for the $M_{int}^{\sigma}$ model. As a matter of fact, the $M_{int}^{\sigma}$ model shows in average the smallest $\Delta_{Err}^{S-m}$ (36%) in comparison to the $M_{int}^{1-EMA}$ (62%) and $M_{int}^{2-EMA}$ (52%) models. This is a clear indication that the interfacial broadening must be considered for an accurate analysis of the optical response of the investigated SLs.

From Table 2, the rising and falling interfacial widths are higher than 2.0 nm for all SLs. Additionally, the Ge EMA at.% is very small (below 10 at.%) for the $M_{int}^{1-EMA}$ model, whereas it is above 50 at.% for the $M_{int}^{2-EMA}$ model. This discrepancy indicates that modelling the $Si_{1-x}Ge_x - Si$ interface as a mixture of two materials having different optical properties is likely an invalid approximation. Next, from the $M_{int}^{\sigma}$ optical model, it is possible to estimate the Ge at.% at the interface, which corresponds to the variable d in equation (2). An average Ge content between 12.0 and 16.0 at.% was estimated for the falling interfacial layer ($Si_{1-x}Ge_x \rightarrow Si$) for all SLs, while for the rising interfacial layer ($Si \rightarrow Si_{1-x}Ge_x$) the average Ge content was found to vary between 11.0 and 17.0 at.%. Furthermore, as shown in **Figure 5(b)**, the interfacial width obtained from the $M_{int}^{\sigma}$ model gives a more reasonable estimation as compared to the EMA-based optical models. **Figure** 5(b) compiles the rising ($Si \rightarrow Si_{1-x}Ge_x$) and falling ($Si_{1-x}Ge_x \to Si$) interfacial widths estimated from APT, EELS, and SE. It is interesting to highlight that the average relative difference between SE and EELS values of the rising and falling interfacial widths are 55% and 39 %, respectively. This relative difference is much larger when SE is compared to APT. However, while the influence of the interface roughness on the optical properties is clearly demonstrated here, it remains very challenging to precisely quantify this interfacial roughness using SE. The latter seems to always overestimate the interfacial broadening as compared to APT and EELS, but only by less than 1 nm. Note that this difference may perhaps come from the fluctuations associated with the lateral scale probed by each method.

In summary, by using Si/SiGe as a model system, APT-generated 3-D maps of SLs and buried interfaces have been employed to quantify the interfacial roughness and the height-height correlation length have been obtained for a variety of sub-10 nm heterostructures. The analysis of iso-concentration maps revealed that the RMS roughness of the buried interfaces is sensitive to the growth temperature with a value varying from ∼0.2 nm (±13.0%) to ∼0.3 nm (±11.5%) in the temperature range of 500-650 ℃. For SLs grown at 500 ℃, the RMS roughness was found to be ∼30% smaller and the horizontal correlation length ∼24% larger, as compared to those grown at 650 ℃. A similar behavior was also observed for the interfacial abruptness, which was found to be practically identical for both $Si \to Si_{1-x}Ge_x$ and $Si_{1-x}Ge_x \to Si$ interface. These studies lay the groundwork to systematically investigate the effects of growth parameters (carrier gas, purging steps between the growth of different layers, different precursors and their partial pressures, material systems, etc.) on the properties of the buried interfaces and their effects on the overall performance. Finally, SE-based optical investigations revealed that an accurate analysis of the optical response of a multilayer heterostructure must take into account the broadening at the interface between different layers.

**Experimental section**

*Growth of the SL samples:* The latter were grown at different temperatures in a reduced pressure chemical vapor deposition (RP-CVD) reactor, on 300 mm undoped Si(001) wafers, using disilane and monogermane as precursors, and hydrogen as carrier gas.

*Cross-sectional transmission electron microscopy:* The sample preparation for APT and XTEM was performed in Dual-FIB microscope, using the standard lamella lift-out technique. The XTEM analysis was conducted in a double cross-section-corrected FEI Titan microscope, operated at 200 kV. The microscope was fitted with Gatan quantum energy filter and a high-brightness electron source. The images were recorded using a high-angle annular dark field detector and the data were processed using the digital micrograph GMS3 software.

*Atom probe tomography:* Prior to the APT tip fabrication in Dual-FIB, a 50 nm thick Ni capping layer was co-deposited on all the samples (using an electron-beam evaporator) in order to protect the top-most part of the samples from ion-implanted damage during the tip fabrication process. APT achieves electric field-induced evaporation of atoms as cations, in a layer-by-layer fashion, from the surface of a needle-like specimen, with the assistance of an ultra-fast pulsed laser.[30,31] In this work, the field evaporation of individual atoms in the APT was assisted by focusing a picosecond pulsed UV laser ($\lambda = 355nm$), with a beam waist smaller than 5µm, on the apex of the needle-shaped specimen. The laser pulse repetition-rate was maintained at 500 kHz throughout. The evaporation rate (ion/pulse) and the pulse energy were varied over a single run. An APT run started with the onset of evaporation of Ni atoms from the capping layer. During this, an evaporation rate of 0.8-1.0 and a laser pulse energy of 30.0 pJ was maintained. As soon as the atoms from the SL appeared at the outer rim of the detector ion map, the evaporation rate was slowed down to 0.2 in a single step and the laser energy was lowered to 4.0-5.0 pJ, in steps of 1.0 pJ. The run was slowed down to ensure that the tip makes a gradual and smooth transition from the Ni cap into the SL without fracturing it. When all the Ni atoms were evaporated and the transition into the SL was complete, the evaporation rate was slowly increased in small steps of 0.20 up to 1.0, ensuring in each step that the automatic voltage ramp is not too steep, a scenario which is well-known to cause tip fracture in APT. After the evaporation made a complete transition from the SL into the Si substrate, the rate was further increased (in steps, reaching up to 3.0-4.0) as well as the pulse energy (in steps, reaching up to 10-15 pJ) in order to collect a substantial amount of substrate atoms as quickly as possible before ending the run.-The base temperature and base pressure within the APT chamber were maintained at 30 K and $3.2 \times 10^{-11}$ Torr, respectively.

*Spectroscopic ellipsometry:* Room-temperature pseudo-dielectric functions $\varepsilon(\omega) = \varepsilon_1(\omega) + i\varepsilon_2(\omega)$ were measured with an automatic rotating analyzer, variable angle spectroscopic ellipsometer (VASE).[32] The samples were mounted and optically aligned with a He-Ne laser in a windowless cell. SE data were collected in the energy range of 0.5-6.0 eV with a 0.01 step size, using multiple angles of incidence (AOI), ranging from 20° to 85°. The Si substrate without HF dip, was also measured by SE under identical conditions to obtain reference data for bulk Si, which compared well with data from Palik[33]. Having a complete and precise structural characterization of the studied SLs is of paramount importance to build an accurate optical ellipsometry model. Indeed, when the optical constants or film structures of a sample are not known well, the ellipsometry results must be cross-correlated with other measurement techniques. To that end, the XTEM- and APT-based structural characterization constitutes a complementary analysis to SE to accurately estimate relevant parameters for the optical model like the periodicity m and the thicknesses of each layer in the SLs.

**Acknowledgments**

This work was supported by NSERC Canada (Discovery, SPG, and CRD Grants), Canada Research Chairs, Canada Foundation for Innovation, Mitacs, PRIMA Québec, and Defence Canada (Innovation for Defence Excellence and Security, IDEaS).

**Conflict of Interest**

The authors declare no competing financial interest.

**Keywords**

silicon-germanium superlattices, atom probe tomography, interfacial roughness, interface correlation length, interfacial width, spectroscopic ellipsometry

**References**

[1] G. Dehlinger, L. Diehl, U. Gennser, H. Sigg, J. Faist, K. Ensslin, D. Grutzmacher, E. Muller, *Science* **2000**, *290*, 2277.
[2] B. M. Maune, M. G. Borselli, B. Huang, T. D. Ladd, P. W. Deelman, K. S. Holabird, A. A. Kiselev, I. Alvarado-Rodriguez, R. S. Ross, A. E. Schmitz, M. Sokolich, C. A. Watson, M. F. Gyure, A. T. Hunter, *Nature* **2012**, *481*, 344.
[3] K.-M. Tan, T.-Y. Liow, R. T. P. Lee, K. M. Hoe, C.-H. Tung, N. Balasubramanian, G. S. Samudra, Y.-C. Yeo, *IEEE Electron Device Lett.* **2007**, *28*, 905.
[4] C. Ciano, M. Virgilio, M. Montanari, L. Persichetti, L. Di Gaspare, M. Ortolani, L. Baldassarre, M. H. Zoellner, O. Skibitzki, G. Scalari, J. Faist, D. J. Paul, M. Scuderi, G. Nicotra, T. Grange, S. Birner, G. Capellini, M. De Seta, *Phys. Rev. Appl.* **2019**, *11*, 014003.
[5] H. Presting, *Thin Solid Films* **1998**, *321*, 186.
[6] S. L. Rommel, T. E. Dillon, M. W. Dashiell, H. Feng, J. Kolodzey, P. R. Berger, P. E. Thompson, K. D. Hobart, R. Lake, A. C. Seabaugh, G. Klimeck, D. K. Blanks, *Appl. Phys. Lett.* **1998**, *73*, 2191.
[7] J. B. Khurgin, *Appl. Phys. Lett.* **2008**, *93*, 091104.
[8] E. Lhuillier, I. Ribet-Mohamed, E. Rosencher, G. Patriarche, A. Buffaz, V. Berger, M. Carras, *Appl. Phys. Lett.* **2010**, *96*, 061111.
[9] M. Franckié, D. O. Winge, J. Wolf, V. Liverini, E. Dupont, V. Trinité, J. Faist, A. Wacker, *Opt. Express* **2015**, *23*, 5201.
[10] K. A. Krivas, D. O. Winge, M. Franckié, A. Wacker, *J. Appl. Phys* **2015**, *118*, 114501.
[11] G. Hellings, H. Mertens, A. Subirats, E. Simoen, T. Schram, L.-A. Ragnarsson, M. Simicic, S.-H. Chen, B. Parvais, D. Boudier, B. Cretu, J. Machillot, V. Pena, S. Sun, N. Yoshida, N. Kim, A. Mocuta, D. Linten, N. Horiguchi, in *2018 IEEE Symp. VLSI Technol.*, IEEE, **2018**, pp. 85–86.
[12] L. Esaki, R. Tsu, *IBM J. Res. Dev.* **1970**, *14*, 61.

[13] S. Koelling, M. Gilbert, J. Goossens, A. Hikavyy, O. Richard, W. Vandervorst, *Appl. Phys. Lett.* **2009**, *95*, 144106.
[14] T. Gredig, E. A. Silverstein, M. P. Byrne, *J. Phys. Conf. Ser.* **2013**, *417*, 012069.
[15] S. Labat, C. Guichet, O. Thomas, B. Gilles, A. Marty, *Appl. Surf. Sci.* **2002**, *188*, 182.
[16] O. C. Hellman, J. A. Vandenbroucke, J. Rüsing, D. Isheim, D. N. Seidman, *Microsc. Microanal.* **2000**, *6*, 437.
[17] O. Dyck, D. N. Leonard, L. F. Edge, C. A. Jackson, E. J. Pritchett, P. W. Deelman, J. D. Poplawsky, *Adv. Mater. Interfaces* **2017**, *4*, 1700622.
[18] D. A. Grützmacher, T. O. Sedgwick, A. Powell, M. Tejwani, S. S. Iyer, J. Cotte, F. Cardone, *Appl. Phys. Lett.* **1993**, *63*, 2531.
[19] E. . Tok, N. . Woods, J. Zhang, *J. Cryst. Growth* **2000**, *209*, 321.
[20] N. Ohtani, S. Mokler, M. H. Xie, J. Zhang, B. A. Joyce, *Jpn. J. Appl. Phys.* **1994**, *33*, 2311.
[21] D. B. Aubertine, N. Ozguven, P. C. McIntyre, S. Brennan, *J. Appl. Phys.* **2003**, *94*, 1557.
[22] D. B. Aubertine, M. A. Mander, N. Ozguven, A. F. Marshall, P. C. McIntyre, J. O. Chu, P. M. Mooney, *J. Appl. Phys.* **2002**, *92*, 5027.
[23] M.-H. Xie, J. Zhang, A. Lees, J. M. Fernandez, B. A. Joyce, *Surf. Sci.* **1996**, *367*, 231.
[24] D. E. Aspnes, *Surf. Sci.* **1980**, *101*, 84.
[25] D. E. Aspnes, *Thin Solid Films* **1982**, *89*, 249.
[26] J. L. Freeouf, *Appl. Phys. Lett.* **1988**, *53*, 2426.
[27] N. V Nguyen, J. G. Pellegrino, P. M. Amirtharaj, D. G. Seiler, S. B. Qadri, *J. Appl. Phys.* **1993**, *73*, 7739.
[28] T. H. Ghong, Y. D. Kim, D. E. Aspnes, M. V. Klein, D. S. Ko, Y. W. Kim, V. Elarde, J. Coleman, *J. Korean Phys. Soc.* **2006**, *48*, 1601.
[29] E. Nolot, J. M. Hartmann, J. Hilfiker, *ECS Trans.* **2014**, *64*, 455.
[30] A. Devaraj, D. E. Perea, J. Liu, L. M. Gordon, T. J. Prosa, P. Parikh, D. R. Diercks, S. Meher, R. P. Kolli, Y. S. Meng, S. Thevuthasan, *Int. Mater. Rev.* **2017**, 1.
[31] S. Mukherjee, H. Watanabe, D. Isheim, D. N. Seidman, O. Moutanabbir, *Nano Lett.* **2016**, *16*, 1335.
[32] D. E. Aspnes, A. A. Studna, *Appl. Opt., AO* **1975**, *14*, 220.
[33] E. D. Palik, *Handbook of Optical Constants of Solids*, Academic Press, London, UK, **1998**.

**Tables and Figures**

Table 1: Composition, layer thickness, and roughness of $(Si)_m/(Si_{1-x}Ge_x)_m$ samples measured with AFM, APT, XTEM, HRXRD, and SE.

| Sample No. | Mean x in $Si_{1-x}Ge_x$ [a] | Mean period thickness in nm [a] from | | | | | | Total SL thickness (XTEM) (nm) | surface rms. roughness (nm) |
|---|---|---|---|---|---|---|---|---|---|
| | | XTEM | | HRXRD | | SE [b] | | | |
| | | $Si_{1-x}Ge_x$ | Si | $Si_{1-x}Ge_x$ | Si | $Si_{1-x}Ge_x$ | Si | | |
| S-3 | 0.31±0.05 | 7.3±0.2 | 6.0±0.2 | 10 | 10 | 8.4±0.9 | 6.5±0.7 | 42.57 | 0.44 |
| S-6 | 0.26±0.05 | 5.8±0.3 | 4.3±0.2 | 5 | 5.5 | 5.7±0.9 | 4.5±1.6 | 57.50 | 0.53 |
| S-12 | 0.25±0.05 | 2.6±0.2 | 2.0±0.1 | 2.2 | 2.7 | 1.2±0.7 | 1.8±0.9 | 50.35 | 0.55 |
| S-16 | 0.25±0.05 | 2.2±0.3 | 1.3±0.2 | 1.5 | 2 | 2.5±0.5 | 1.6±0.8 | 56.25 | 0.64 |

(a) The mean includes the last $Si_{1-x}Ge_x$ layer which had grown thicker and with higher Ge content than the remaining $Si_{1-x}Ge_x$ layers. The Ge content was estimated from APT measurements and not HRXRD.
(b) The $M_{int}^{\sigma}$ optical model was used to estimate the average thicknesses of the $Si_{1-x}Ge_x$ and Si layers in the SLs.

Table 2: Quantitative comparison of all the proposed optical models ($M_X$, $M_{int}^{1-EMA}$, $M_{int}^{2-EMA}$, and $M_{int}^{\sigma}$) where the first column shows the behavior of the average thickness relative error $\Delta_{Err}^{S-m}$ (%) as a function of the studied superlattices ($S-m$) and the optical model used. The second and third column present the extracted $Si_{1-x}Ge_x \rightarrow Si$ and rising $Si \rightarrow Si_{1-x}Ge_x$ interfacial thickness for the interfacial optical models. The values in parenthesis represents the error associated with the values.

| Sample \ Model | $\Delta_{Err}^{S-m}$ (%) | | | | $Si \rightarrow Si_{1-x}Ge_x$ Interfacial width (nm) | | | $Si_{1-x}Ge_x \rightarrow Si$ Interfacial width (nm) | | |
|---|---|---|---|---|---|---|---|---|---|---|
| | $M_X$ | $M_{int}^{1-EMA}$ | $M_{int}^{2-EMA}$ | $M_{int}^{\sigma}$ | $M_{int}^{1-EMA}$ | $M_{int}^{2-EMA}$ | $M_{int}^{\sigma}$ | $M_{int}^{1-EMA}$ | $M_{int}^{2-EMA}$ | $M_{int}^{\sigma}$ |
| S-3 | 41(2) | 39(4) | 40(1) | 37(1) | 2.6(1.6) | 2.57(1.10) | 1.48(0.50) | 2.6(1.8) | 2.35(0.85) | 1.79(0.40) |
| S-6 | 69(3) | 43(3) | 43(6) | 22(1) | 2.05(0.78) | 2.37(0.57) | 1.64(0.40) | 2.74(0.70) | 1.23(0.50) | 1.86(0.46) |
| S-12 | 72(4) | 77(3) | 51(3) | 48(1) | 2.26(0.95) | 2.84(0.50) | 1.45(0.35) | 2.40(0.85) | 2.80(0.40) | 1.35(0.45) |
| S-16 | 93(5) | 85(2) | 75(3) | 32(2) | 2.35(0.83) | 2.62(0.45) | 1.35(0.35) | 2.54(0.35) | 2.20(0.40) | 1.26(0.40) |

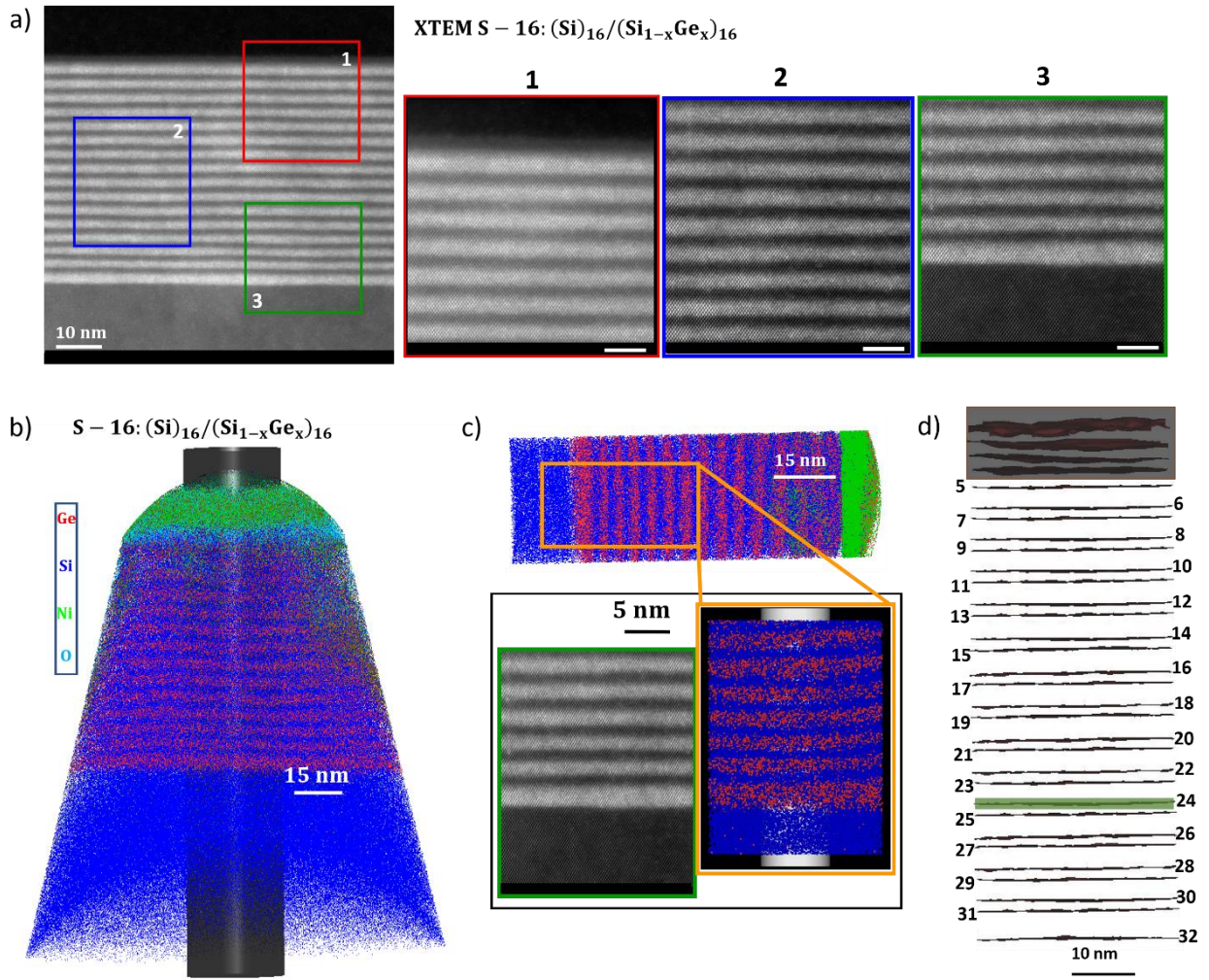

**Figure 1.** XTEM and APT 3-D reconstruction of sample S-16. (a) XTEM image of sample S-16. A zoom into the selected regions of the image is shown alongside using the numbered boxes. All scale bars for the images within these boxes correspond to 5 nm. (b) 3-D atom-by-atom APT reconstruction of sample S-16. (c) Top: 3-D reconstruction of the atoms lying within the black cylinder (diameter 30 nm) at the center of the 3D reconstruction shown in (b). Bottom: Comparison demonstrating the correspondence of the XTEM image with the APT reconstruction, by zooming into the last (bottom-most) 6-7 periods of sample S-16. (d) Si/SiGe heterointerfaces of sample S-16, represented as Ge iso-concentration surfaces drawn within the reconstruction shown in (c)-top.

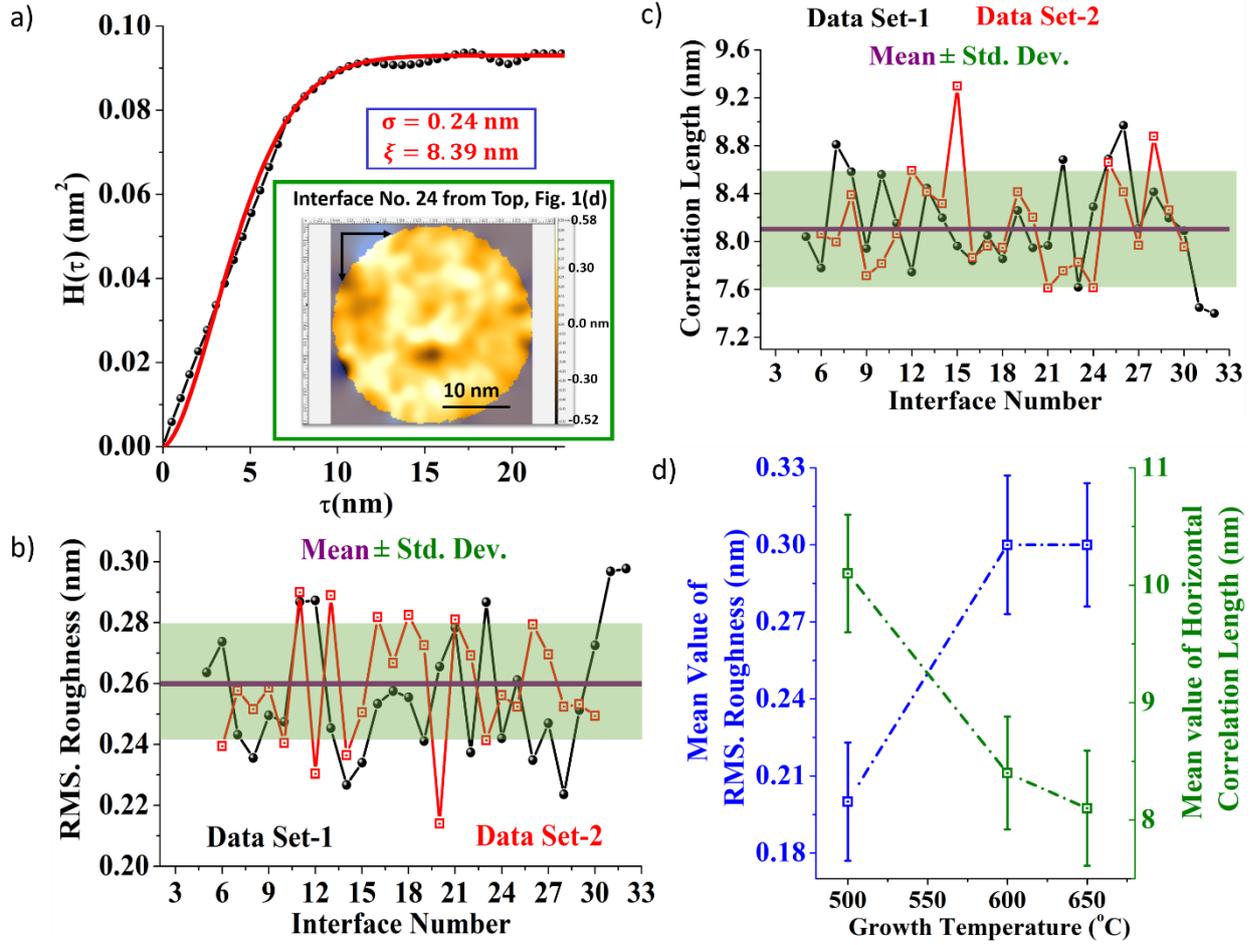

**Figure 2.** Extraction of the RMS. roughness and horizontal correlation length. (a) The evolution of height-height correlation function with horizontal length of interface 24 in **Figure** 1(d) and the corresponding fitting function (red). The color-coded height distribution of the same interface is shown in the inset. For convenience of display, some of the vales of the color bar have been highlighted. The extracted value of RMS roughness (σ) and the horizontal correlation length (ξ) for interface 9 of sample S-16 are highlighted in the inset. The accuracy of the fit or the $R^2$ value was found to be better than 0.99. (b) The extracted values of σ of all the interfaces of sample S-16 over two different data sets. For data set-2 of sample S-16, the top 6 interfaces were damaged from Ga ion implantation and the tip fractured while transiting through the last $Si_{1-x}Ge_x$ layer. Hence, the data for interface number 7 to 30 are shown. (c) The extracted values of ξ of all the interfaces of sample S-16 over two different data sets. In (b) and (c), the mean value of σ and ξ is shown using the thick purple line and the uncertainty (shown using the green transparent box) represent the standard deviation of the data. (d) Evolution of σ (in blue) and ξ (in green) of the SLs as a function of the growth temperature. The dotted lines are for guide to the eye. The error bars represent the standard deviation over all the interfaces and over different data sets of a sample.

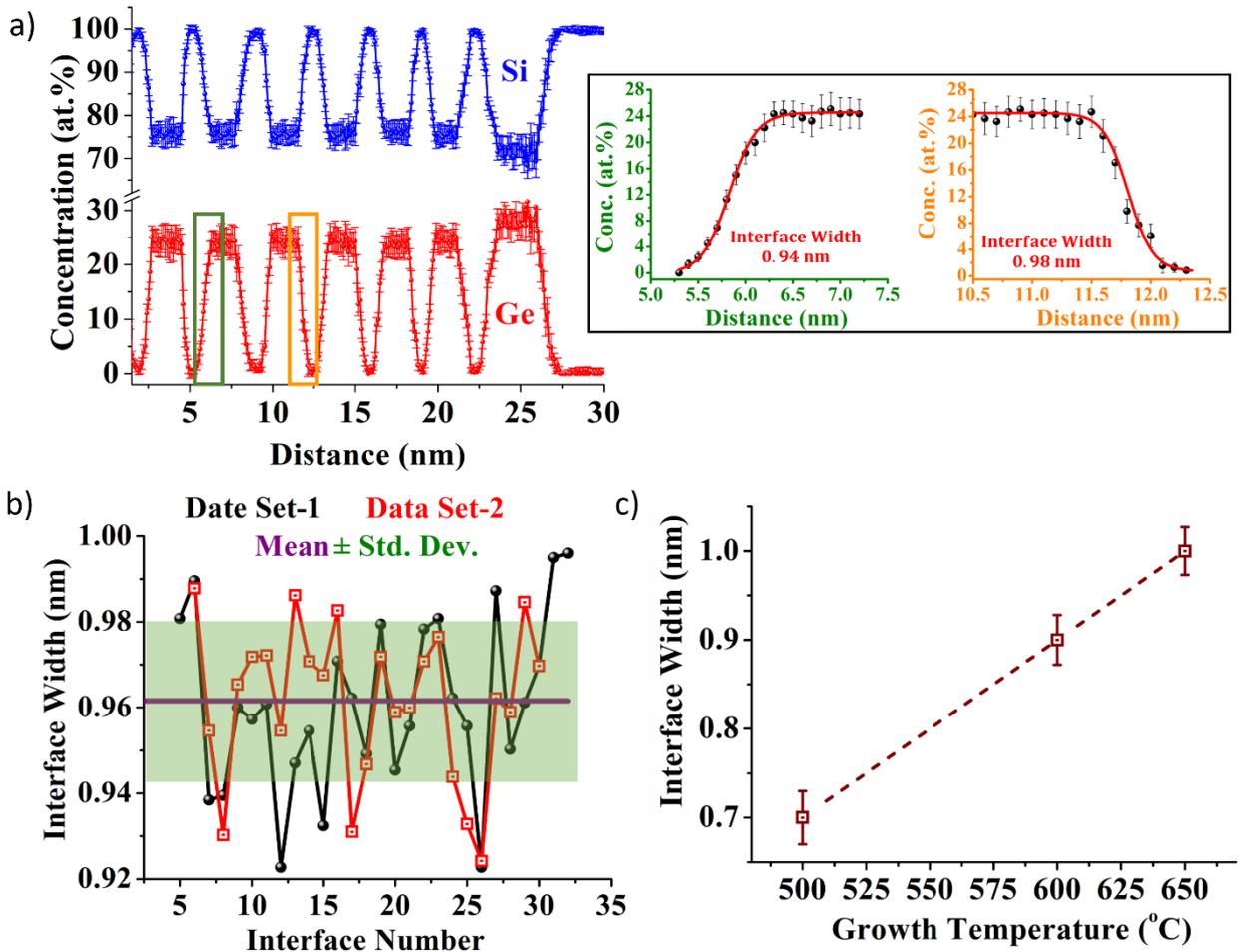

**Figure 3**. Extraction of interfacial width. (a) Concentration profiles of Si and Ge of the last seven periods of sample S-16, recorded at a fixed bin width of 0.2 nm (from within the white cylinder of diameter 8 nm, within the APT reconstruction shown in **Figure** 1(c): bottom). The interfaces marked by the orange and the green rectangle are used in the inset demonstrate the extraction of the interfacial width. Inset: (Left) A zoom in at the green rectangle at the 12[th] interface from bottom, showing a rising Ge concentration and corresponding sigmoid fit (in red). The extracted value of the interface width is 0.95 nm. (Right) A zoom in at the 9[th] interface from bottom, showing a falling Ge concentration and corresponding sigmoid fit (in red). The extracted value of the interface width is 0.96 nm. The accuracy of the fit or the $R^2$ value for both the plots in the inset was found to be better than 0.99. (b) The interface width of all the interfaces of sample S-16 over two different data sets. The mean value of the interface width is shown using the thick purple line and the uncertainty (shown using the green transparent box) represent the standard deviation of the data. (c) Evolution of the mean interface width of the SLs as a function of the growth temperature. The dotted line is a guide to the eye. The error bars represent the standard deviation over all the interfaces and over different data sets of a sample.

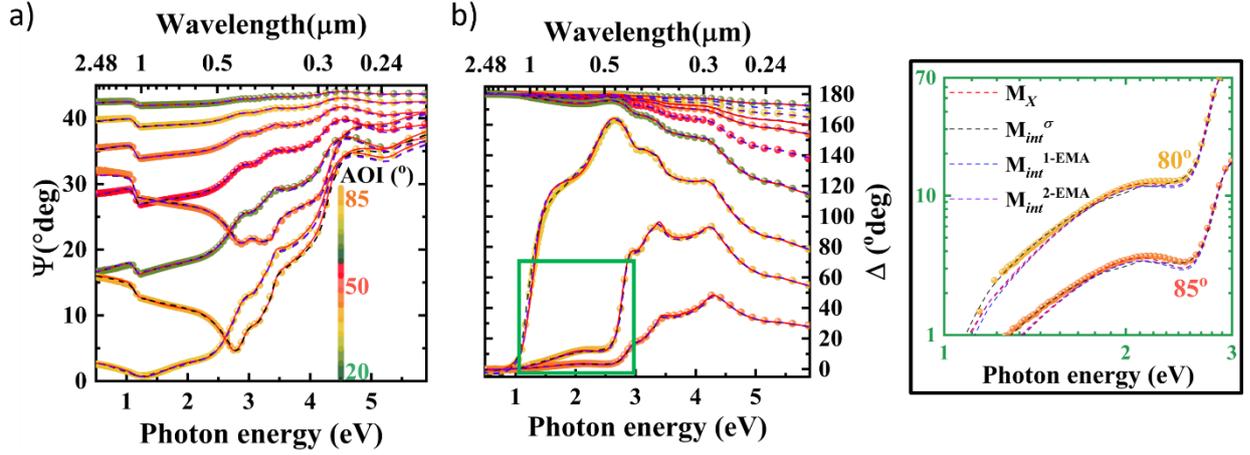

**Figure 4**. Spectroscopic ellipsometry-measured parameters for sample S-6. The experimental parameters (a) Ψ and (b) Δ, extracted from the SE measurements. The fit to the experimental data based on the four optical models mentioned in the text (see the details of the optical models in the SI). The red line is the fitted spectra obtained with $M_X$ model, whereas the black, blue, and purple dashed-lines are respectively the result of the $M_{int}^{\sigma}$, $M_{int}^{1-EMA}$, and $M_{int}^{2-EMA}$ optical models. The spectral region from 1 to 3 eV marked by the green rectangle in (b) is used in the inset (within the black rectangular box) to highlight qualitatively the accuracy of the different optical models by plotting the parameter Δ in a loglog scale.

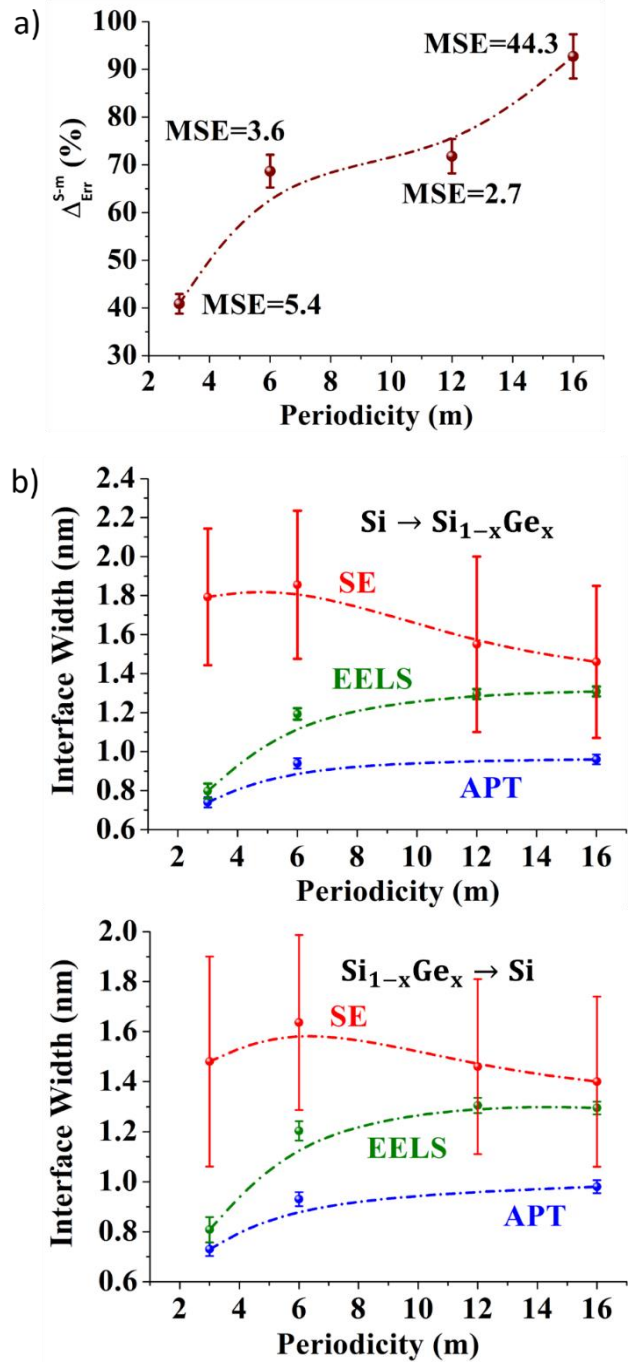

**Figure 5**. Comparison of the MSE and interface width across different SL samples measured using EELS, APT, and SE. (a) The variation of the average thickness relative error $\Delta_{Err}^{S-m}$ (%) as a function of the number of periods m evaluated from the $M_X$ optical model. The MSE is also overlaid in the figure to show that even though the MSE is small which is usually indicative of accurate SE model, the error of the developed optical model is still high, thus justifying the necessity to use $\Delta_{Err}^{S-m}$ as a metric to quantify the accuracy of the optical model, in addition to the MSE. (d) A comparison of the variation of the rising (Si → $Si_{1-x}Ge_x$) and falling ($Si_{1-x}Ge_x$ → Si) interfacial width estimated from three different characterization techniques: red from SE, olive-green from EELS and blue from APT. The error bar estimation for the SE data is the standard deviation evaluated from the iterative process and the dashed lines are cubic

B-spline fit to the data (the fits do not serve any physical purpose and are simply for guidance to the eye).

# Supporting Information

**Section 1: Additional details of APT**

Note that the top few Si and $Si_{1-x}Ge_x$ layers in the APT reconstruction in **Figure** 1(b) of the main manuscript appears to be artificially intermixed with the Ni atoms from the capping layer. This feature showed up in all the APT reconstruction of all the samples investigated in this work. In all likelihood, the microbalance controller in the electron-beam evaporator gave a faulty reading and the Ni layer was much thinner than the expected 50 nm. As a result, towards the end of the tip fabrication when the tips were being polished, it is likely that the energetic $Ga^+$ ions were able to pass through this thin Ni layer and get implanted at the top part of the SL, resulting in this artificial intermixing of the top 7-8 nm of all the samples. The layers that happen to fall within this 7-8 nm region from the top, are consequently left out in our analysis.

An APT reconstruction always makes an implicit assumption that the surface wherefrom the atoms are being evaporated (as cations) is hemispherical. However, when a tip makes a transition from one material to another with different evaporation fields ($F_{ev}$), the radius of curvature of the tip ($r_{tip}$) changes locally to accommodate the difference in their $F_{ev}$. A material with higher $F_{ev}$ requires a smaller radius of curvature compared to a material of lower $F_{ev}$ in order to maintain the same evaporation rate, according to the relation $F_{ev} = V_{bias}/kr_{tip}$, where k is a constant related to the shape of the tip and $V_{bias}$ is the dc voltage applied to the tip. This phenomenon is known to cause artificial expansion and compression at the interface of two materials with different $F_{ev}$, the artificial expansion occurring while making a transition from a material with lower $F_{ev}$ to a material with higher $F_{ev}$ while the compression happening during the reverse transition. If the z-axis is taken to be the axis of the tip, then the artificial compression and expansion at the interface of two different materials causes an artificial increase and decrease in the atomic density respectively, at the center of the 3D distribution. A z-redistribution algorithm was proposed by Vurpillot et *al.* (details in reference[1]) that relaxes the atomic density in the z-direction, thereby correcting this artifact. Some improvements in interfacial widths were also reported after performing the density-corrected 3D reconstruction. However, for the samples investigated in this work no difference in the interface widths were found for the 3D APT reconstructions performed with or without the need for density correction.[2,3]

**Section 2: APT, XTEM and EELS – Additional Data**

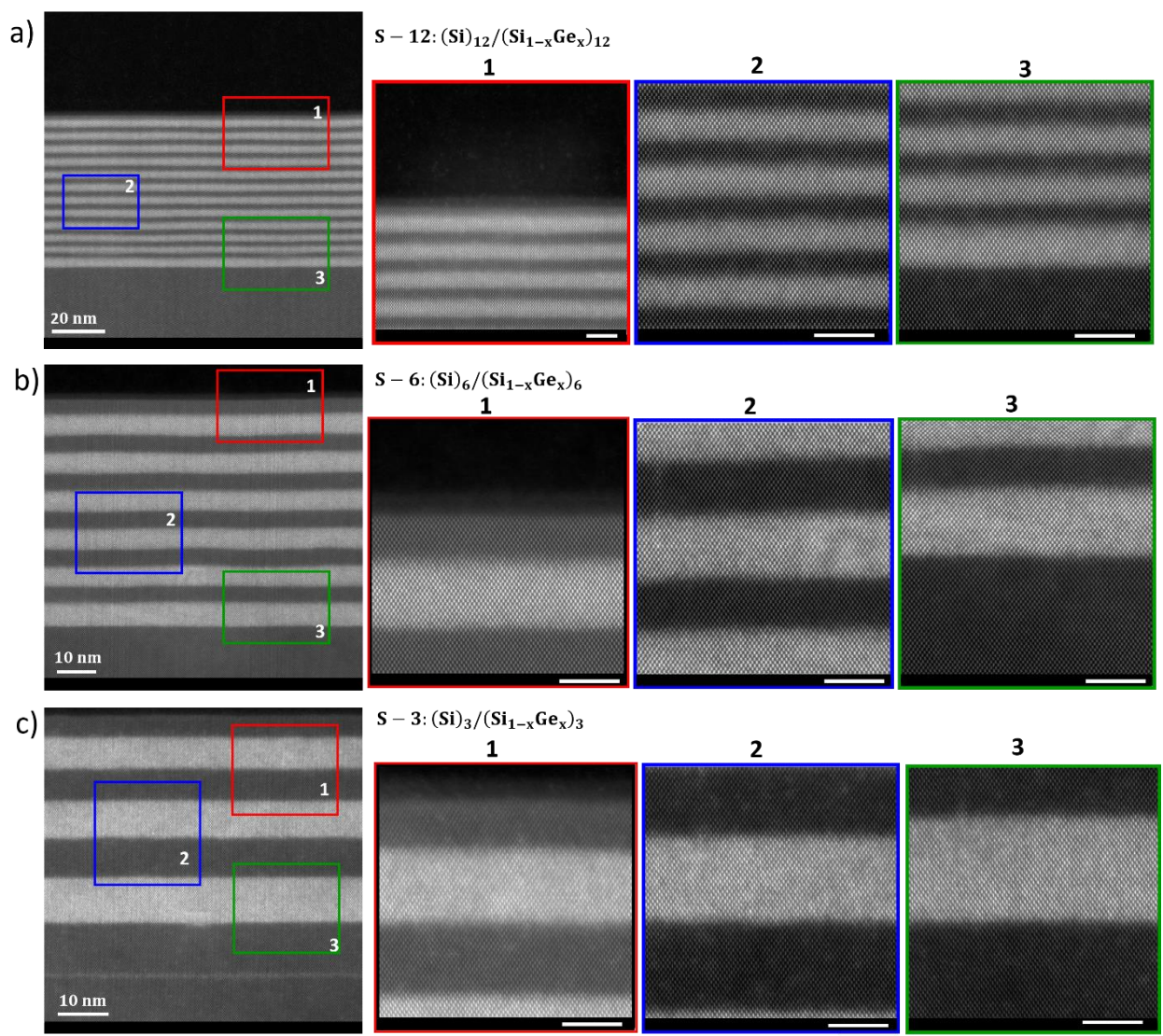

**Figure S1**. (a) XTEM image of sample (a) S-12, (b) S-6, and (c) S-3. A zoom into the selected regions of a particular sample is shown using the numbered color-coded rectangular boxes. All scale bars for the images within the numbered color-coded rectangular boxes are 5 nm.

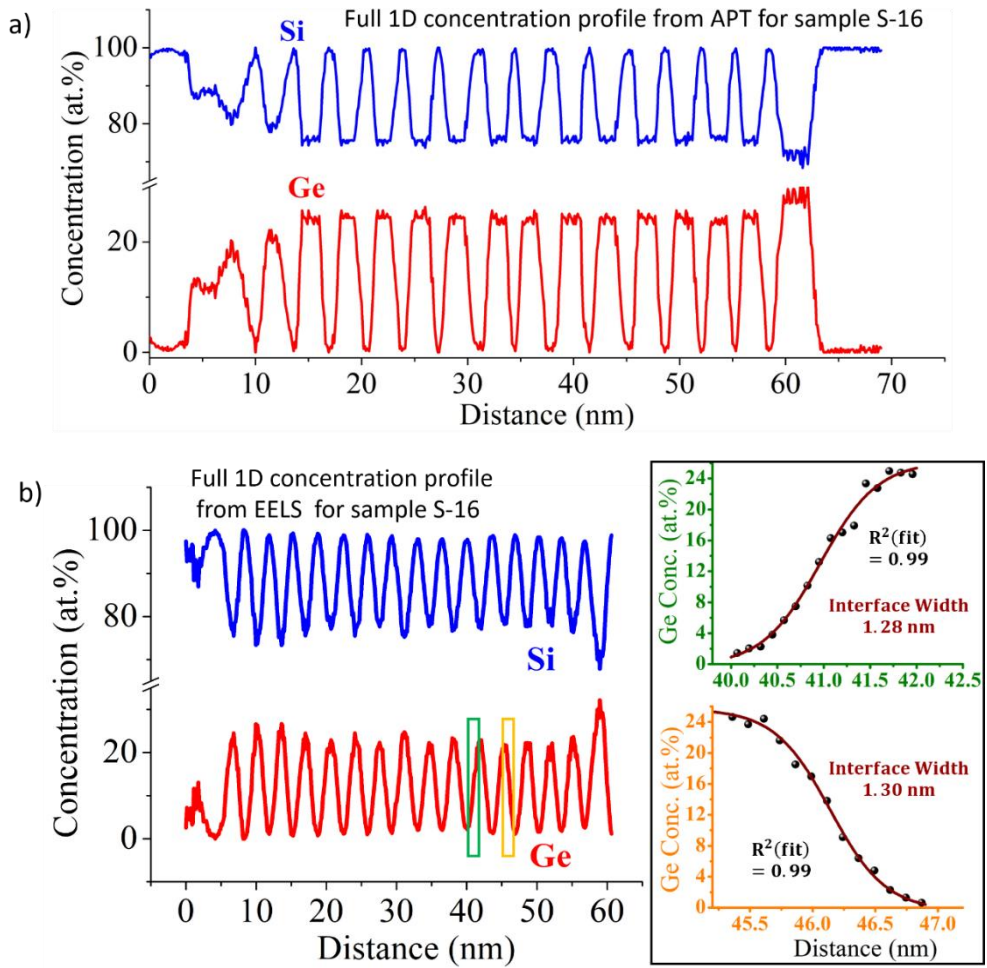

**Figure S2**. Full 1D concentration profile of sample S-16, extracted from (a) APT and (b) EELS. Inset: The zoom in onto the Ge concentration across the 9[th] interface (in orange) and the 12[th] interface (in olive-green) from the bottom and the corresponding sigmoid fit (in wine-brown). The extracted values of the interface widths and the goodness of the fit ($R^2$) have been highlighted within each graph. These are the same two interfaces shown in the inset of **Figure** 3(a) of the main manuscript. Notice the broadening of the concentration profile and larger values of the interface width that is extracted from EELS data, as compared to the APT.

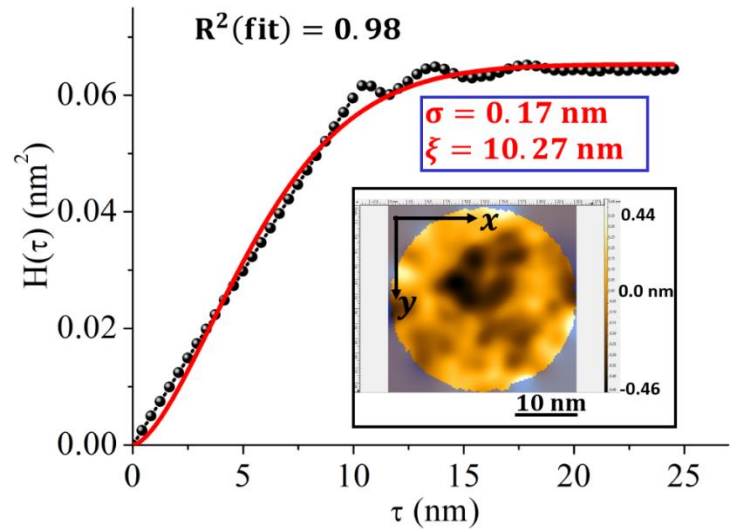

**Figure S3**. An example of the evolution of the height-height correlation function H(τ) as a function of the correlation length (τ) of sample S-3. The chosen interface is the 3$^{rd}$ from top of sample S3, marked by the black rectangle in the Ge concertation profile in **Figure** S7(a). The extracted value of RMS. roughness (σ), correlation length (ξ), and the goodness of the fit ($R^2$) for this interface are also shown. Inset: The color-coded height distribution (z) image of the interface. The scale bar represents 10 nm. For convenience of display, the ends (+0.44 nm, -0.46 nm) and the middle (0.00 nm) of the color bar are marked.

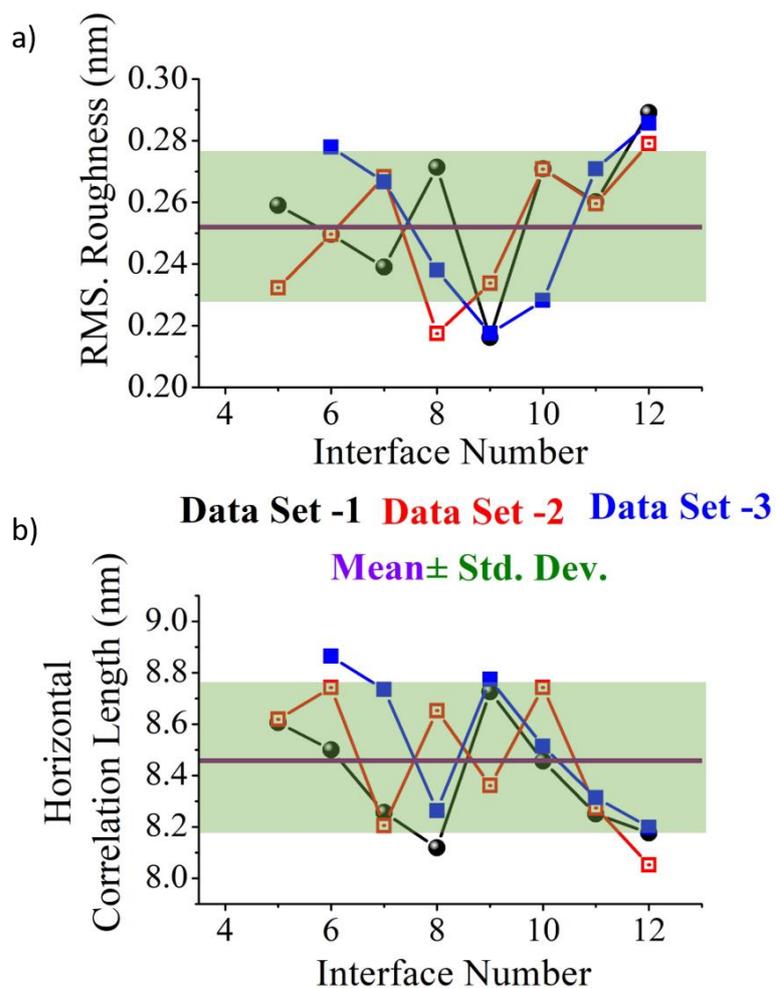

**Figure S4**. The values of the (a) RMS. roughness ($\sigma$) and (b) horizontal correlation length ($\xi$) of different interfaces and over different data sets of the sample S-6. The mean values denoted by the thick purple line in the respective graphs and the uncertainty (shown using the green transparent box) represent the standard deviation of the data.

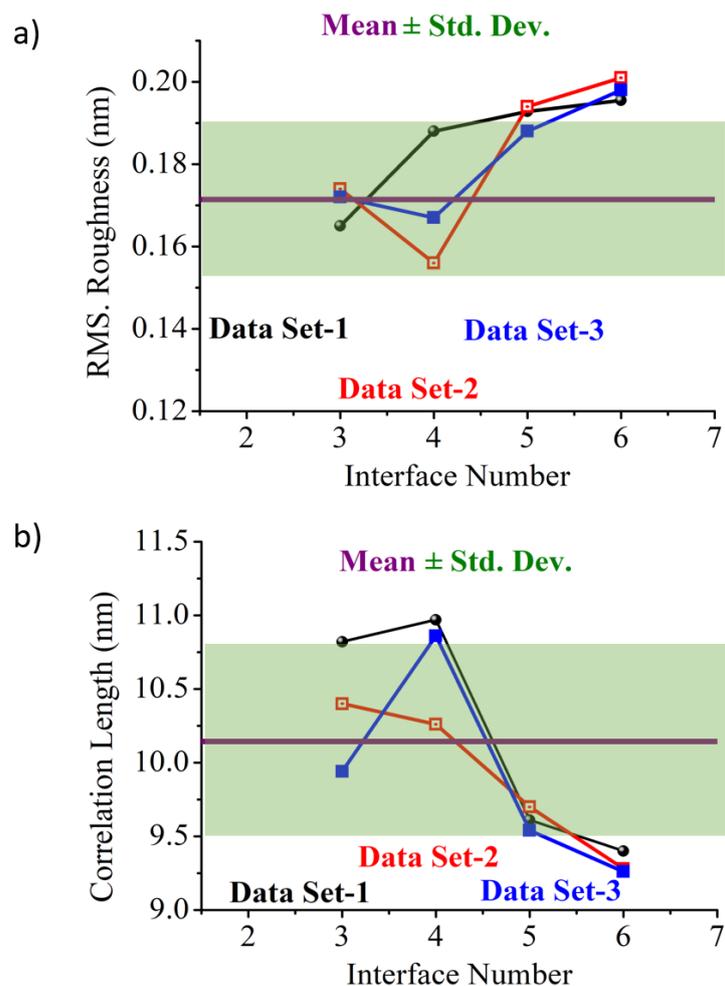

**Figure S5**. The values of the (a) RMS. roughness ($\sigma$) and (b) horizontal correlation length ($\xi$) of different interfaces and over different data sets of the sample S-3. The mean values denoted by the thick purple line in the respective graphs and the uncertainty (shown using the green transparent box) represent the standard deviation of the data.

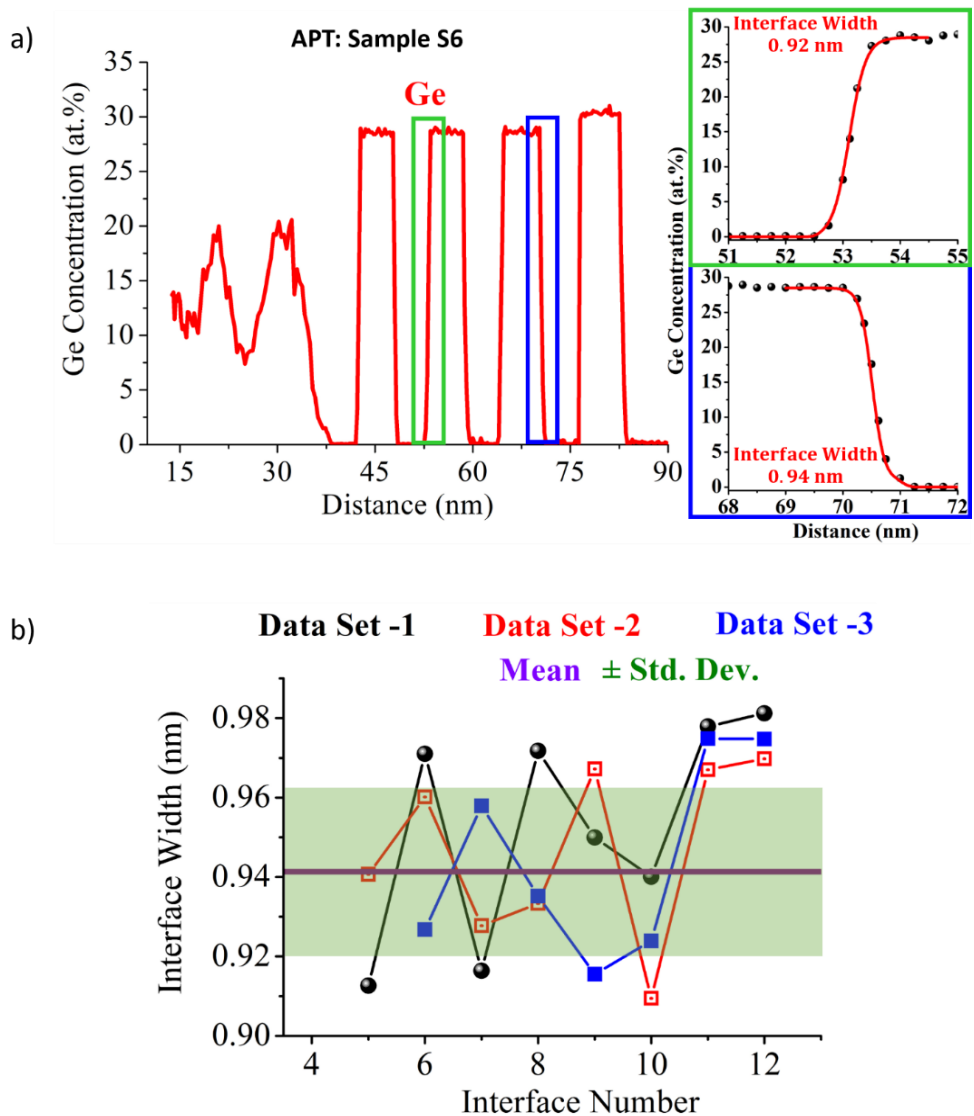

**Figure S6**. (a) 1D Ge concentration profile (at a fixed bin width of 0.2 nm) extracted from APT reconstruction of sample S-6. Inset: Rising (within the green box, zoomed in at the green rectangle in the full profile) and falling (within the blue box, zoomed in at the blue rectangle in the full profile) Ge concentration profile along with the sigmoid fit (red curve) and the extracted value of the interface width. (b) The values of the interface width of different interfaces and over different data sets of the sample S-6. The mean values denoted by the thick purple line and the uncertainty (shown using the green transparent box) represent the standard deviation of the data.

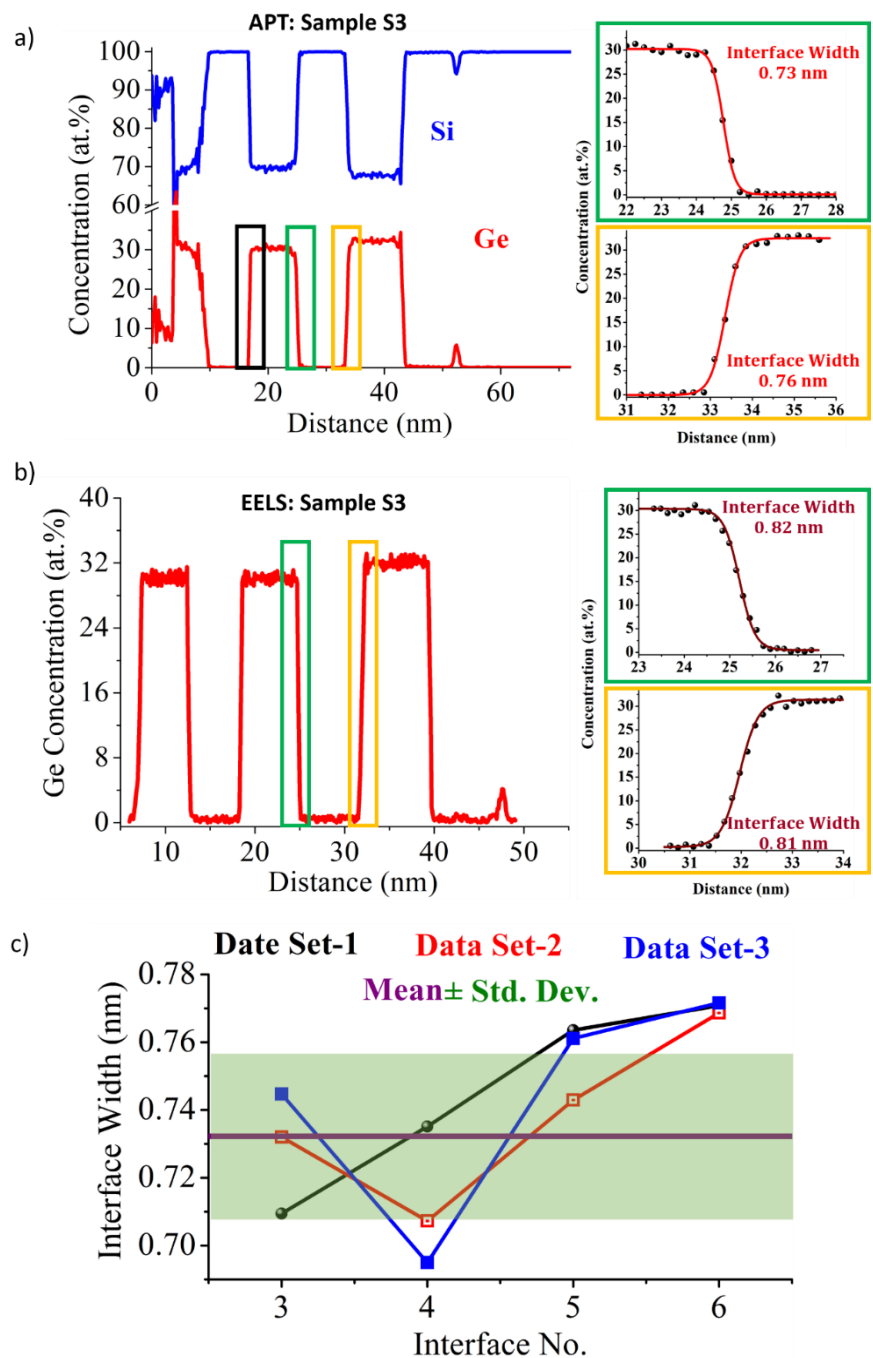

**Figure S7**. (a) 1D concentration profile (at a fixed bin width of 0.2 nm) extracted from APT reconstruction of sample S-3. Inset: falling (within the green box, zoomed in at the green rectangle in the full profile) and rising (within the orange box, zoomed in at the orange rectangle in the full profile) Ge concentration profile along with the sigmoid fit (red curve) and the extracted value of the interface width. (b) 1D Ge concentration profile of sample S3 from EELS data Inset: Rising and falling Ge concentration profile (at the same two interfaces as that for the APT profile in (a)) along with the sigmoid fit (red curve) and the extracted value of the interface width. (c) The extracted values of interface width of different interfaces and over different data sets of the sample S-3. The mean is denoted by the thick purple lines and the uncertainty (shown using the green transparent box) represent the standard deviation of the data.

**Section 3: Details and additional data related to spectroscopic ellipsometry**

**3.1. A few basic concepts**: Ellipsometry Spectroscopic (SE) ellipsometry is a non-destructive optical method for characterizing both thin and thick layers of either amorphous or crystalline material. This technique is well suited for measuring films with thicknesses of a few angstroms up to several microns, particularly for those materials that have well established refraction indices values. The SE uses electromagnetic wavelengths from an extreme ultra-violet to the far infrared. The variable angle spectroscopic ellipsometry (VASE) uses change in the state of polarization of light upon reflection for characterization of surfaces, interfaces, and thin films. The analysis of the VASE experiment in hetero-structured $Si_{1-x}Ge_x$/Si multilayer systems can enable the assessment of the critical point (CP) energies as well as information on interfacial thickness, crystallinity, roughness and composition of individual layers. In ellipsometry, the measured ratio $\rho$ of the reflection coefficient $r_p$ and $r_s$ can be expressed in terms of the amplitude ratio $\tan \Psi$ and the phase angle $\Delta$:[4]

$$\rho = \frac{r_p}{r_s} = \tan \Psi \, e^{-i\Delta} \qquad (E1)$$

where the two ellipsometry parameters $\Psi$ and $\Delta$ can be obtained directly from the SE measurements; $r_p$ and $r_s$ represent parallel and perpendicular reflection coefficients to the plane of incidence, respectively. The complex pseudo-dielectric function $\varepsilon(\omega) = \varepsilon_1(\omega) + i\varepsilon_2(\omega)$ can be derived from SE data by using a three-phase model, and is defined as follows

$$\varepsilon(\omega) = \sin^2 \varphi + \sin^2 \varphi \tan^2 \varphi \left[\frac{1-\rho}{1+\rho}\right]^2 \qquad (E2)$$

where $\varphi$ is the angle of incidence (AOI). For each sample, the SE measurements were undertaken for energy ranging between 0.5 eV and 6 eV with a step size of 0.01 eV at several AOI $\varphi$ to increase the accuracy of the subsequent analysis. The non-focused spot size has a diameter of around 2 mm, and so a projected major axis between 4.7 and 7.7 mm (for angles of 65° and 75°, respectively). While $\varepsilon(\omega)$ is the "true" function for bulk material in the absence of an overlayer, it is called the "pseudodielectric" function for multilayer (including just bulk plus overlayer) systems. In such cases, the genuine values can only be obtained using multiphase models.[5] Once the model has been built, one varies the physical parameters using the Levenberg-Marquardt algorithm to minimize the mean-squared error (MSE) defined by

$$\text{MSE} = \sqrt{\frac{1}{3N-M} \sum_{i=1}^{3N} \left|\frac{\rho_i^{mod} - \rho_i^{exp}}{\Delta \rho_i^{exp}}\right|^2} \qquad (E3)$$

where N is the number of data points (all photon energies, incidence angles, and samples), M the number of parameters, $\rho_i^{exp}$ the three experimental quantities (ellipsometric angles $\Psi$ and $\Delta$ and depolarization) at each data point, $\rho_i^{mod}$ the quantities calculated from the model, and $\Delta \rho_i^{exp}$ the experimental errors. The MSE is a figure of merit for how well the present set of parameters fits the ellipsometric data. However, to accurately estimate the interface thickness, an angle dependant iteration procedure needs to be undertaken.

**3.2. Details related to the optical model:** To avoid the observer-bias, input related to interface width, layer thickness, and composition from XTEM and APT was not been used to develop the interface-related optical model. Rather, the information from these complementary structural characterization techniques have only been used to quantitatively verify the accuracy of the built model. The layer thicknesses and Ge composition in each sublayer of the SL, obtained from HRXRD, was then used as initial estimate for the SE optical model. The relative difference of the mean period thickness estimation between XTEM and SE was evaluated to be around 26% in average for all the studied samples which is a qualitative indication of the goodness of the proposed optical model. The surface roughness was measured by AFM (see **Figure** S10 in section 4, for sample S-3) which demonstrates the top surface to be smooth and devoid of any microroughness or large voids. Another indication of the surface smoothness and the sample homogeneity is that none of the samples exhibited depolarization (<1%) for the energy range studied here. It is quite well-known that when the optical constants or film structures of a sample are not known well, the ellipsometry results must be justified using other characterization methods. In general, there are two boundary conditions that can be applied when setting the initial iteration and for confirming the physical validity of the converged value after iterations. First, the thickness should not be lower than the thickness of a one-unit cell of the materials. For films, this is the thinnest physical limit for the layer-by-layer deposition, while for interfaces this takes into consideration any interface roughness effects. Second, the thickness should not be higher than five times the photon penetration depth, D, since beyond this limit the material can considered to be bulk-like from an optical point of view.[5] A rigorous treatment of the photon penetration depth inside the $(Si)_m/(Si_{1-x}Ge_x)_m$ SL has been elaborated in section 3.4. The effective penetration depth for all the samples is approximately 1-50 nm above 2.5 eV, whereas it can reach up to 30 µm for photon energy below 2.5 eV, which is more than enough to cover layer thicknesses of 2-10 nm and an interfacial width of

~1-2 nm. To accurately model the interface thickness in the SL, special care needs to be given to the ellipsometric measurement to guarantee that the optical fitting process will converge to a physical solution. Briefly, an incident-angle dependant iteration was used where the number of AOI is closely related to the number of interfaces in the SLs.[6] A detailed explanation of the measurement procedure is presented in section 3.3. Starting from the HRXRD initial estimation of the thicknesses of the Si and $Si_{1-x}Ge_x$ layers, as well as the Ge content in each SL with a ±20% variance, the experimentally obtained SE data was fitted to all the optical models described below.

**Figure** S8(a) presents a layer-by-layer optical model (named $M_X$) where the interfaces are omitted, whereas **Figure** S8(b) highlights the incorporation of the interfacial layers (this model has been labeled $M_{int}$). In the $M_{int}$ model, the initial layers thicknesses of the Si layer ($d_{Si}^{(i)}$) and that of the $Si_{1-x}Ge_x$ layer ($d_{SiGe}^{(i)}$) are taken from HRXRD. Three different optical models are proposed to build the dielectric constant of the interface layer as shown in the inset of the **Figure** S8(b). First, the Bruggeman effective medium approximation (B-EMA) is used,[7–11] which mixes the dielectric functions of two adjacent layers to form that of the interface ($M_{int}^{1-EMA}$). Second, another variant of the first model is used ($M_{int}^{2-EMA}$). The main difference between the $M_{int}^{1-EMA}$ and $M_{int}^{2-EMA}$ models is the materials chosen to form the dielectric constant of the interface. The former mixes the dielectric constants of Si and Ge, whereas the latter uses the dielectric constant of $Si_{1-x}Ge_x$ and Si. Third, a new parametric graded interfacial alloy model ($M_{int}^{\sigma}$) is introduced where the Si content inside the interface layer is graded and described by equation (2) of the main manuscript, and the floating parameters are the scaling parameter $d$ and the interfacial sharpness $\mathcal{L}$. The interface width $d_{int}^{(i)}$ is coupled to $\mathcal{L}$ via $d_{int}^{(i)} = 4\mathcal{L}$. Grading is achieved by dividing the thin film into a series of $\mathcal{N}$ sublayers that have varying optical properties to approximate the index gradient profile. It is important to note that because ellipsometric models always consider optically homogeneous layers[4,5] and any layer gradient described by a mathematical function must also be discretized into homogeneous sublayers. However, choosing arbitrarily large numbers of sublayers can be computationally demanding. To find the optimal number of sublayers $\mathcal{N}$, the mean squared error (MSE) change was investigated with respect to $\mathcal{N}$. Due to the thin interface layer, the MSE was found not to improve for $\mathcal{N}$ bigger than 21. Thus, for the $M_{int}^{\sigma}$ model, $\mathcal{N}$ was fixed to 21. For the $M_{int}^{1-EMA}$ and $M_{int}^{2-EMA}$ models, starting with the assumption that a physical mixture of bulk material (Si and Ge for $M_{int}^{1-EMA}$ or Si and $Si_{1-x}Ge_x$ for $M_{int}^{2-EMA}$), finding the thickness as well as the EMA % in the interface was made possible. For instance, in the $M_{int}^{1-EMA}$ model, EMA % represents the Ge content within the host material Si. The floating parameters in this case become the thickness $d_{int}^{(i)}$ and the EMA %. An iterative process was developed to extract the desired set parameters. The first iteration involves varying the layer thicknesses ($d_{SiGe}^{(i)}$ and $d_{Si}^{(i)}$) and keeping fixed the already set interface width ($d_{int}^{(i)}$). The second iteration fixes the layer thicknesses and varies the interface width. This iterative process is repeated until the MSE between each consecutive step is smaller than a set tolerance value of $10^{-3}$ and the gradient of $\Delta_{Err}^{S-m}$ (defined in the following section, see equation (3) of the main manuscript) is minimized.

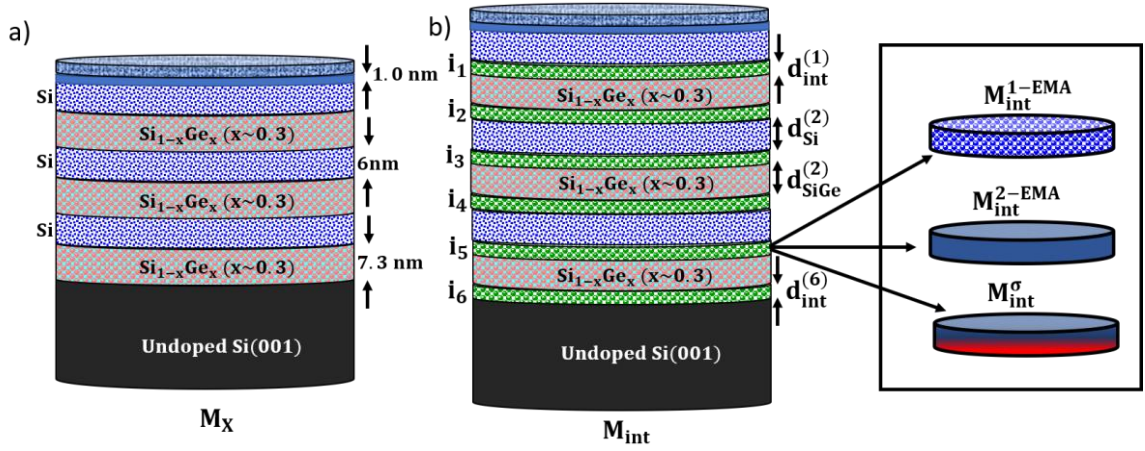

**Figure S8**. The different spectroscopic ellipsometric models used throughout this work where (a) $M_X$ is the XTEM-based SE optical model where the layer-by-layer structure is the building block of the model. The thicknesses and the composition of Ge in each layer are evaluated. The schematic also incorporates a $SiO_2$ layer as well as a surface roughness layer (top layer), which are kept fixed during the optical modeling. (b) By adding an interfacial layer $i_m$ between each Si and $Si_{1-x}Ge_x$ layers of the $M_X$ model, with thickness $d_{int}^{(i)}$, the interface can be analyzed in all the SLs. Inset: Three different methods are proposed to physically model the interface. The first $M_{int}^{1-EMA}$ consists of using a Bruggeman-EMA model where a physical mixture between Si and Ge is assumed, whereas in $M_{int}^{2-EMA}$ the interface is composed of a physical mixture between Si and $Si_{1-x}Ge_x$. Finally, a graded interfacial layer is also studied in the $M_{int}^{\sigma}$ model where the Si content in the interface is determined by a sigmoid function as described in equation (2) of the main manuscript.

**3.3. Incident-angle dependant iteration procedure:** The theory describing wave propagation in a stratified medium is well established in literature.[4,5,12] It allows accurate evaluation for the reflection coefficient of a single thin film or a multilayered structure on a substrate. Developing the exact expression for the reflection coefficient is out of the scope of this work. However, for a multilayered system with N layers ($L_N$ is the $N^{th}$ layer), the obtained reflection coefficient, $r_{amb,SL}$ and from equation (E1), $\rho$, $\Psi$, and $\Delta$ depend on the dielectric constant of each layer in the SL and the substrate, along with the thickness of each layer and the angle of incidence

$$\rho = \rho_{theory}\left(\varepsilon_{L_1}^1, \ldots \varepsilon_{L_N}^N, \varepsilon_{subs}, d_{L_1}^1, \ldots d_{L_N}^N, \varphi\right)$$
$$= \tan\Psi_{meas}\, e^{-i\Delta_{meas}} \tag{E4}$$

However, a detailed analysis is required to separate the different contribution of each layer. A versatile method to extract and separate the dielectric constant of each individual layer of the superlattice is discussed. If there are multiple layers of thin films composed of different materials, and the $\varepsilon(\omega)$ of each material is unknown or different from their bulk forms (as in the case of the current SLs), then the analysis becomes more complicated. This is due to the increased numbers of unknowns but only one equation (equation E4), which hinders a straightforward mathematical solution. To overcome this problem, it can be noted that the light phase, $\delta = 2\pi d/\lambda\, n \cos\varphi$ depends mainly on two parameters: the incident angle, $\varphi$ and the layer thickness, d. Consequently, by varying either of the previously mentioned parameters, the number of equations can match the number of the unknown variables (in the case of the S-3 sample, six which corresponds to the six interfaces thicknesses $d_{inter}^{(i)}, i = 1, \ldots 6$ in S-3). On the one hand, varying the thickness of the layers means at least six superlattices with same Ge content (in each layer) and same periodicity (m = 3) and different layers thicknesses need to be fabricated. This can be achieved with a proper growth control, but it is still a challenging task. In the other hand, varying the AOI $\varphi$ is the better choice. The $\Psi$ and $\Delta$ measurement need to be done at 6 or more different incident angles (55°, 60°, 65° 70°, 75° and 80°) for the S-3 samples. Table T1 presents the different AOI considered for each sample to diversify the equation (E4)

Table T1: AOI for the measurement of Ψ and Δ for the different SLs. The total number of the AOI need to be equal or bigger than the number of the interfaces.

| Sample | Number of interfaces | AOI | step |
|---|---|---|---|
| S-3 | 6 | 55°-80° | 5° |
| S-6 | 12 | 30°-85° | 5° |
| S-12 | 24 | 25°-83° | 2.5° |
| S-16 | 32 | 20°-83° | 1° |

**3.4. Photon Penetration Depth inside a superlattice:** Photon penetration depth, D, is a measure of how deeply light can penetrate a medium. It is defined as the depth at which the intensity of the radiation inside the medium falls to 1/e of its initial value, $I_0$, where e is the natural constant. The penetration depth, D, can be obtained from the dielectric constant $\varepsilon(\omega) = \varepsilon_1(\omega) + i\varepsilon_2(\omega)$ of the material according to[12]

$$D = \frac{\lambda\sqrt{\varepsilon_1}}{2\pi\varepsilon_2} \qquad (E5)$$

If the material is multilayered like the $(Si_{1-x}Ge_x)_m/(Si)_m$ superlattice, the intensity drop depends on the penetration depth of each constituent layer, in this case the Si layer, the GeSi layer, the interface layer, and the Si substrate,

$$I(z) = I_0 \exp\left(-\left[\sum_{i=1}^{m} \frac{d_{Si}^{(i)}}{D_{Si}^{tfilm}} + \sum_{i=1}^{m} \frac{d_{Si_{1-x}Ge_x}^{(i)}}{D_{Si_{1-x}Ge_x}^{tfilm}} + \sum_{i=1}^{2m} \frac{d_{inter}^{(i)}}{D_{inter}}\right.\right.$$
$$\left.\left.+ \frac{z - \sum_{i=1}^{m} d_{Si}^{(i)} - \sum_{i=1}^{m} d_{Si_{1-x}Ge_x}^{(i)} - \sum_{i=1}^{2m} d_{inter}^{(i)}}{D_{Si}^{Bulk}}\right]\right)$$
(E6)

where $z \geq \left(\sum_{i=1}^{m} d_{Si}^{(i)} - \sum_{i=1}^{m} d_{Si_{1-x}Ge_x}^{(i)} - \sum_{i=1}^{2m} d_{inter}^{(i)}\right)$ is along the direction perpendicular to and measured from the surface of the superlattice, $d_{Si}^{(i)}, d_{Si_{1-x}Ge_x}^{(i)}$, and $d_{inter}^{(i)}$ are respectively the thicknesses of the Si, the $Si_{1-x}Ge_x$ and the interface layers. From equation (E6), the effective penetration depth, $D_{eff}$, of the $(Si_{1-x}Ge_x)_m/(Si)_m$ superlattice can be expressed as

$$D_{eff} = D_{Si}^{Bulk} + \sum_{i=1}^{m}\left[d_{Si}^{(i)}\left(1 - \frac{D_{Si}^{Bulk}}{D_{Si}^{tfilm}}\right) + d_{Si_{1-x}Ge_x}^{(i)}\left(1 - \frac{D_{Si}^{Bulk}}{D_{Si_{1-x}Ge_x}^{tfilm}}\right)\right]$$
$$+ \sum_{i=1}^{2m} d_{inter}^{(i)}\left(1 - \frac{D_{Si}^{Bulk}}{D_{inter}}\right)$$
(E7)

**3.5. 12 nm thin-film crystalline Silicon optical properties:** Modifying the thickness of Si layer has a direct affect on the optical properties, mainly the complex dielectric constant. To that end, we measured an SOI sample with SE for different incident angle varying from 60° to 85° with a 1° step. **Figure** S9(a) shows the ellipsometry parameter Δ as well as the built optical model with an MSE of 1.12. The SOI sample is composed of a Si substrate, on top of which a SiO$_2$ layer of 15 nm is deposited and it is then capped with a 12 nm c-Si, as shown in the inset of **Figure** S9(b). The Si layers inside the studied SLs has an average thickness of 3.6±2.5 nm where the error represents the standard deviation of the Si thickness of all the SLs. Thus, approximating the Si dielectric constant within the SLs with that of a 12 nm layer, is a stronger approximation than using its bulk counterparts. In **Figure** S9(b), a comparison between the bulk c-Si[13–15] and the 12 nm thin-film c-Si is shown, where the black and green crossed circles is the dielectric constant of the thin-film c-Si. From **Figure** S9(b), the difference in the dielectric constant is noticeable for the energy range between 3.46 and 4.21 eV, which corresponds to the $E_1$, $E_1 + \Delta_1$, and $E_2$ interband transitions in Si.

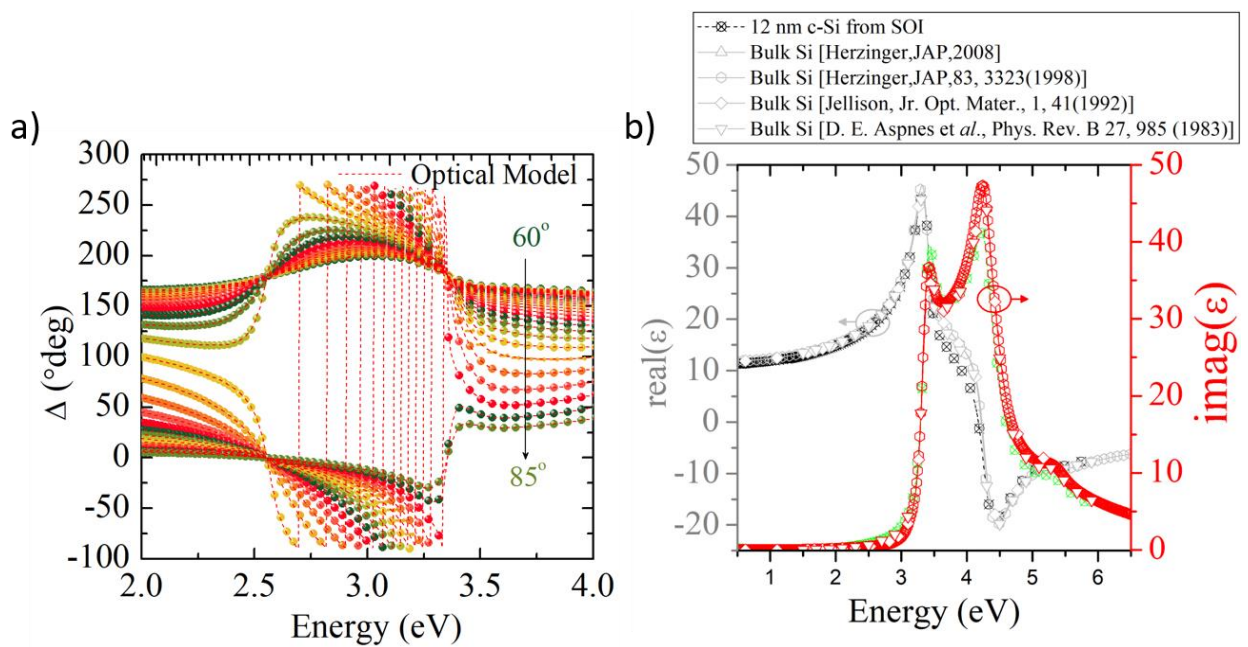

**Figure S9**. (a) Spectroscopic ellipsometric angle $\Delta$ for the SOI sample for an AOI between 60° and 85° with a 1° step. (b) A comparison of the dielectric constant of the c-Si layer extracted from the SOI sample to that of a bulk Si from different research groups.

**Section 4: AFM Characterization**

Quantifying the morphology of the top surface of the $(Si_{1-x}Ge_x)_3/(Si)_3$ gives insight into the growth quality of the superlattice, and the dislocation distribution in the sample. **Figure** S10 shows the root mean square (RMS) surface roughness of the S-3 sample.

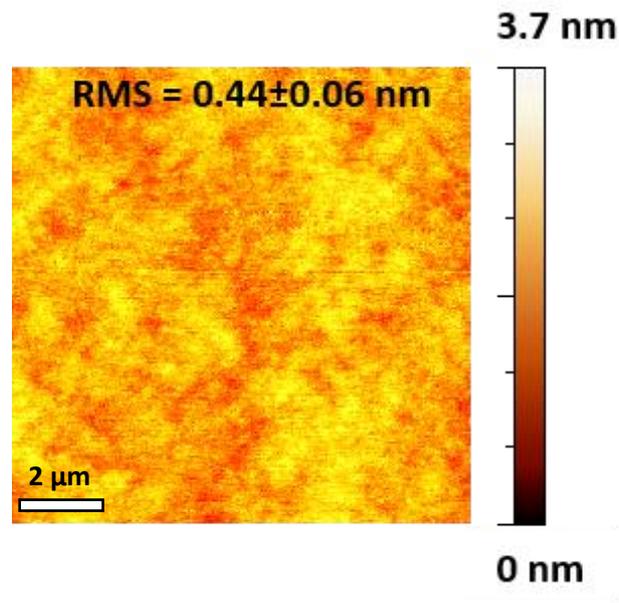

**Figure S10**. Root Mean Square (RMS) surface roughness for the $(Si_{1-x}Ge_x)_3/(Si)_3$ SL sample, measured with AFM with a 10 µm × 10 µm images.

**References**

[1] F. Vurpillot, D. Larson, A. Cerezo, *Surf. Interface Anal.* **2004**, *36*, 552.
[2] O. Dyck, D. N. Leonard, L. F. Edge, C. A. Jackson, E. J. Pritchett, P. W. Deelman, J. D. Poplawsky, *Adv. Mater. Interfaces* **2017**, *4*, 1700622.
[3] J. G. Brons, A. A. Herzing, K. T. Henry, I. M. Anderson, G. B. Thompson, *Thin Solid Films* **2014**, *551*, 61.
[4] H. G. Tompkins, E. A. Irene, I. An, H. Arwin, C. Chen, R. W. Collins, A. S. Ferlauto, J. N. Hilfiker, J. Humlícek, J. Gerald E. Jellison, J. Lee, F. A. Modine, A. Röseler, M. Schubert, J. A. Zapien, *Handbooks of Ellipsometry*, William Andrew Publishing And Springer-Verlag GmbH & Co. KG, **2005**.
[5] H. Fujiwara, *Spectroscopic Ellipsometry Principles and Applications*, John Wiley & Sons, Ltd, **2003**.
[6] T. C. Asmara, I. Santoso, A. Rusydi, *Rev. Sci. Instrum.* **2014**, *85*, 123116.
[7] D. E. Aspnes, *Surf. Sci.* **1980**, *101*, 84.
[8] D. E. Aspnes, *Thin Solid Films* **1982**, *89*, 249.
[9] J. L. Freeouf, *Appl. Phys. Lett.* **1988**, *53*, 2426.

[10]  N. V Nguyen, J. G. Pellegrino, P. M. Amirtharaj, D. G. Seiler, S. B. Qadri, *J. Appl. Phys.* **1993**, *73*, 7739.
[11]  T. H. Ghong, Y. D. Kim, D. E. Aspnes, M. V. Klein, D. S. Ko, Y. W. Kim, V. Elarde, J. Coleman, *J. Korean Phys. Soc.* **2006**, *48*, 1601.
[12]  M. Born, E. Wolf, *Principles of Optics*, Cambridge University Press, Cambridge MA, **1999**.
[13]  D. Aspnes, A. Studna, *Phys. Rev. B* **1983**, *27*, 985.
[14]  G. E. J. Jr., *Opt. Mater. (Amst).* **1992**, *1*, 41.
[15]  C. M. Herzinger, B. Johs, W. A. McGahan, J. A. Woollam, W. Paulson, *J. Appl. Phys.* **1998**, *83*, 3323.